\documentclass[epsfig]{article}
\usepackage{amssymb}

\usepackage{graphicx}
\usepackage{amsmath}

\input epsf.sty
\newtheorem{theorem}{Theorem}
\newtheorem{acknowledgement}[theorem]{Acknowledgement}

\input{tcilatex}
\makeatletter \@addtoreset{equation}{section}
\renewcommand{\theequation}{\thesection.\arabic{equation}}

\begin{document}

\title{\rightline{\mbox {\normalsize {Lab/UFR-HEP/0305/GNPHE/0306}}}\vspace{1cm}%
\textbf{NC Geometry and Fractional Branes}}
\author{El Hassan Saidi\thanks{%
E-mail: h-saidi@fsr.ac.ma} \\
{\small Lab/UFR-High Energy Physics, Physics Department,\ Faculty of
Science, Rabat, Morocco}\textit{.}\\
{\small and} \\
{\small National Grouping in High Energy Physics, Focal point: Faculty of
Science, Rabat, Morocco}}
\maketitle
\date{}

\begin{abstract}
Considering complex $n$-dimension Calabi-Yau homogeneous hyper-surfaces $%
\mathcal{H}_{n}$ with discrete torsion and using Berenstein and Leigh
algebraic geometry method, we study Fractional D-branes that result from
stringy resolution of singularities. We first develop the method introduced
in hep-th/0105229 and then build the non commutative (NC) geometries for
orbifolds $\mathcal{O}=\mathcal{H}_{n}/\mathbf{Z}_{n+2}^{n}$ with a discrete
torsion matrix $t_{ab}=exp[{\frac{i2\pi }{n+2}}{(\eta _{ab}-\eta _{ba})}]$,
\ $\eta _{ab}\ \in SL(n,\mathbf{Z})$. We show that the NC manifolds $%
\mathcal{O}^{(nc)}$ are given by the algebra of functions on the real $%
(2n+4) $ Fuzzy torus $\mathcal{T}_{\beta _{ij}}^{2(n+2)}$ with deformation
parameters $\beta _{ij}=exp{\frac{i2\pi }{n+2}}{[(\eta _{ab}^{-1}-\eta
_{ba}^{-1})}\ q_{i}^{a}\ q_{j}^{b}]$ with $q_{i}^{a}$'s being charges of $%
\mathbf{Z}_{n+2}^{n}$. We also give graphic rules to represent $\mathcal{O}%
^{(nc)}$ by quiver diagrams which become completely reducible at orbifold
singularities. It is also shown that regular points in these NC geometries
are represented by polygons with $(n+2)$ vertices linked by $(n+2)$ edges
while singular ones are given by $(n+2)$ non connected loops. We study the
various singular spaces of quintic orbifolds and analyze the varieties of
fractional $D$ branes at singularities as well as the spectrum of massless
fields. Explicit solutions for the NC quintic $\mathcal{Q}^{(nc)}$ are
derived with details and general results for complex $n$ dimension orbifolds
with discrete torsion are presented. \newline
\end{abstract}

\tableofcontents

\newpage

\newpage

\section{Introduction}

Recently some interest has been given to build non commutative (NC)
extensions of Calabi-Yau orbifolds $\mathcal{O}$ with discrete torsion \cite
{1}-\cite{10}. These NC manifolds, which are involved in the twisted sector
of type II string compactification, go beyond the standard non commutative $%
\mathbf{R}_{\theta }^{n}$ and NC$\ \mathcal{T}_{\theta }^{k}$\ torii
examples considered in brane physics \cite{11}-\cite{16}. They offer a
manner to resolve non geometric stringy singularities \cite{17}, present a
natural framework to study fractional $D$ branes\cite{18}-\cite{23} and
provide insight into the relationship between discrete torsion and the
T-dual B field \cite{46}-\cite{48,49}. The basic idea of this construction
may be summarized as follows: the usual closed string compact target
subspace $\mathcal{O}\sim \mathcal{M}/G$, with $G\subset \mathbf{Z}_{m}^{k}$%
, is viewed as the commuting central subalgebra $\mathcal{Z}\left( \mathcal{A%
}^{nc}\right) $ of a NC algebra $\mathcal{A}^{nc}$\ representing $\mathcal{O}%
^{nc}$. The latter can be thought of as a fibration $\mathbf{B}\bowtie
\QTR{sl}{F}$ with a base $\mathbf{B=}\mathcal{O}$ \ and as a fiber $\QTR{sl}{%
F}=\mathcal{A}_{G}$ given by the algebra of the group of discrete torsion $G$%
. The centre $\mathcal{Z}\left( \mathcal{A}^{nc}\right) $ deals with the
closed string sector, which in $\alpha ^{\prime }$ leading order contains
massless type II supergravity; while $\mathcal{A}^{nc}/\mathcal{Z}\left(
\mathcal{A}^{nc}\right) \sim \QTR{sl}{F}$ is associated with open strings
describing Supersymmetric Yang-Mills theories.

From the algebraic geometry point of view,\ the NC manifold \ $\mathcal{O}%
^{nc}$ \ is covered by a finite set of holomorphic matrix coordinate patches
$\mathcal{U}_{(\alpha )}=\{Z_{i}^{(\alpha )};1\leq i\leq n\;\alpha
=1,2,\ldots \}$ \ and holomorphic transition functions mapping \ $\mathcal{U}%
_{(\alpha )}$\ \ to \ $\mathcal{U}_{(\beta )}$; $\phi _{(\alpha ,\beta )}$:
\ $\mathcal{U}_{(\alpha )}\ \ \rightarrow \ \mathcal{U}_{(\beta )}$. In
other words $\mathcal{A}^{nc}$ is covered by a collection of NC local
algebras $\mathcal{A}^{nc}\mathcal{{_{(\alpha )}}}$, with analytic maps $%
\phi _{(\alpha ,\beta )}$ on how to glue $\mathcal{A}_{(\alpha )}^{nc}$ and $%
\mathcal{A}^{nc}\mathcal{{_{(\beta )}}}$. These $\mathcal{A}^{nc}\mathcal{{%
_{(\alpha )}}}$ local algebras have centers $\mathcal{Z}_{\left( \alpha
\right) }=\mathcal{Z}\left( \mathcal{A}_{(\alpha )}^{nc}\right) $; \ when
glued together give precisely the commutative target space manifold $%
\mathcal{O}$. In this way a commutative singularity of $\mathcal{O}\simeq
\mathcal{Z}\left( \mathcal{A}^{nc}\right) $ can be made smooth in the NC
space $\mathcal{A}^{nc}$ \cite{1,2}. This idea was successfully used to
build NC ALE spaces and some realizations of Calabi-Yau threefolds (CY3)
such as the quintic threefolds $\mathcal{Q}$. In this regards it was shown
that the NC quintic $\mathcal{Q}^{nc}$ extending $\mathcal{Q}$,\ when
expressed in the coordinate patch $Z_{5}=I_{id}$, \ is given by the
following special algebra,
\begin{eqnarray}
Z_{1}Z_{2} &=&\alpha \ Z_{2}Z_{1},\qquad Z_{3}Z_{4}=\beta \gamma \
Z_{4}Z_{3},\ \ \ \ \ \ \ \ \ \ \ (a)  \notag \\
Z_{1}Z_{4} &=&\beta ^{-1}\ Z_{4}Z_{1},\qquad Z_{2}Z_{3}=\alpha \gamma \
Z_{3}Z_{2},\ \ \ \ \ \ \ \ (b)  \notag \\
Z_{2}Z_{4} &=&\gamma ^{-1}\ Z_{4}Z_{2},\qquad Z_{1}Z_{3}=\alpha ^{-1}\beta \
Z_{3}Z_{1},\ \ \ \ \ (c) \\
Z_{i}Z_{5} &=&\ Z_{5}Z_{i},\qquad i=1,2,3,4;  \notag
\end{eqnarray}
where $\alpha ,\beta $\ and\ $\gamma $\ are fifth roots of the unity, the
parameters of the $\mathbf{Z}_{5}^{3}$\ discrete group. The $Z_{i}$\ 's are
the generators of $\mathcal{A}^{nc}$. One of the main features of this non
commutative algebra is that its centre $\mathcal{Z(Q}^{nc})$ coincides
exactly with $\mathcal{Q}$, the commutative quintic threefolds. In \cite{2},
a special solution for this algebra using $5\times 5$\ matrices has been
obtained and in \cite{24} a class of solutions for eqs(1.1), depending on
the orbifold group charge vectors, has been worked out and some partial
results regarding higher dimensional CY hypersurfaces were given.

In this paper, we study other aspects of NCCY orbifolds with discrete
torsion completing the partial results obtained in \cite{2,24}-\cite{17};
but also explore new issues such the connection between NC geometry and
fractional D branes or the link between discrete torsion and rational tori .
More precisely we will compute the explicit dependence of NCCY orbifolds in
the discrete torsion of the orbifold group and study the varieties of the
fractional $D$ branes at singularities. We show that the Berenstein and
Leigh (BL) construction is just a NC torus fibration with base the CY
orbifold in complete agreement with the idea of emergent dimension developed
recently in \cite{46}, see also \cite{47}. The result concerning NC torus
fibration can be directly seen on eqs (1.1) which, for the case of generic
complex $d$ dimension CY hypersurfaces, can be rewritten formally as $%
Z_{i}Z_{j}=\alpha _{ij}\ Z_{j}Z_{i}$. A careful inspection of the solution
of these eqs, shows that they describe\ a fibration whose base is indeed the
$\mathcal{H}_{d}$ commutative space and a fiber given by a NC rational torus
defined as,

\begin{equation}
\alpha _{ij}^{n+2}=1,\qquad Z_{i}^{n+2}\sim I_{id}.
\end{equation}
Throughout this paper, we will also show that the origin of these rational
torii is due to a nice property of $Z_{n+2}$ orbifolds, which induce a NC
torus fibration on $\mathcal{H}_{d}$. In particular, we show that the NC
structure has two sources: (i) either induced by quantum symmetries as usual
or (ii) by considering NC complex one cycles as in the CDS solution for
Matrix model compactification of M theory\cite{11}. Both solutions lead to a
finite number of fractional $D$ branes and provide a new way to think about
the B field.

The organization of this paper is as follows: In section $2$, we review some
general features of the commutative quintic and develop its non commutative
extensions using the constrained method. We also complete partial results
obtained in literature. The analysis we will develop in this section applies
as well to all CY hypersurfaces and moreover to CY manifolds embedded in
toric varieties \cite{33}. In section $3$, we develop non commutative
geometry induced by discrete torsion and work out the full class of torsion
dependent solutions for NC quintic. In section $4$, we use this result to
derive general solutions for the NCCY orbifolds. This analysis recovers the
results of [17] as special cases. We also study the symmetries of the moduli
space of NC solutions we have obtained and discuss the varieties of
fractional $D$ branes as well as the full spectrum of massless fields on the
$D$ branes. In section $5$, we collect some general results on NC complex $d$%
-dimensional CY orbifolds with a discrete torsion matrix $t_{ab}=exp[{\frac{%
i2\pi }{d+2}}{(\eta _{ab}-\eta _{ba})}]$, \ $\eta _{ab}\ \in SL(d,\mathbf{Z}%
) $ and study the various classes of varieties of fractional D branes at
singularities. The NC manifolds are given by the algebra of functions on the
Fuzzy torus $\mathcal{T}_{\beta _{ab}}^{2d}$, where $\beta _{ab}=exp{\frac{%
i2\pi }{d+2}}{[(\eta _{cd}^{-1}-\eta _{dc}^{-1})}\ q_{a}^{c}\ q_{b}^{d}]$ .
In section $6$, we give our conclusion.

\section{NC Quintic $\mathcal{Q}^{nc}$}

In this section, we study the algebraic geometry approach for building NC
quintic by using constraint eqs method. This analysis applies as well to any
complex $d$ dimension CY homogeneous hypersurface $\mathcal{P}_{d+2}{%
(z_{1},\dots ,z_{d+2})=0}$ embedded in $\mathbf{CP}^{d+1}$. Among the
results we will derive here, we prove that the NC quintic obtained by
Berenstein and Leigh is a special torus fibration based on $\mathcal{Q}$. We
show that orbifolds of the quintic are characterized by a SL$\left( 3,%
\mathbf{Z}\right) $ matrix $\eta _{ab}$ whose antisymmetric part encodes
discrete torsions of the $Z_{5}^{3}$ orbifold group.

\subsection{The Quintic $\mathcal{Q}$}

To begin consider the complex analytic homogeneous hypersurface $\mathcal{P}%
_{5}{(z_{1},\dots ,z_{5})}$ given by,

\begin{equation}
z_{1}^{5}+z_{2}^{5}+z_{3}^{5}+z_{4}^{5}+z_{5}^{5}+a_{0}%
\prod_{i=1}^{5}z_{i}=0,  \label{20}
\end{equation}
where $(z_{1},z_{2},z_{3},z_{4},z_{5})$ are commuting homogeneous
coordinates of $\mathbf{CP}^{4}$, the complex four dimensional projective
space, and $a_{0}$ a non zero complex moduli parameter. This polynomial
describes a well known CY3s namely the commutative quintic denoted in this
paper by $\mathcal{Q}$. Besides invariance under permutations of the $z_{i}$
variables, $\mathcal{Q}$ has a set of geometric discrete isometries acting
on the homogeneous coordinates $z_{i}$ as:

\begin{equation}
z_{i}\rightarrow z_{i}\omega ^{q_{i}^{a}},  \label{21}
\end{equation}
with $\omega ^{5}=1$ and $q_{i}^{a}$ integer charges to which we will refer
here below to as the CY charges\footnote{%
By CY charges we intend the $\mathbf{q}^{a}=\left( q_{i}^{a}\right) $
vectors that define a basis for phase symmetries $\mathbf{Z}_{d+2}^{d}$ of
the complex $d$ dimension CY hypersurface $z_{1}^{d+2}+...+z_{d+2}^{d+2}+%
\prod_{i=1}^{d+2}z_{i}=0$. Since the \textit{a-th} $\mathbf{Z}_{d+2}$ factor
\ acts on the $z_{i}$ local variables as $z_{i}^{\prime }=z_{i}\omega
^{q_{i}^{a}}$ , invariance of \ the above polynomial requires that $\omega
=\exp i\frac{2\pi }{d+2}$ and moreover $\sum_{i=1}^{d+2}q_{i}^{a}=0$. This $%
q_{i}^{a}$ constraint relation is known to be equivalent to the vanishing
condition of the first Chern class of CY manifolds.}. These are the entries
of the following 5d vectors $\mathbf{q^{1}}=\mathbf{(}%
q_{1}^{1},q_{2}^{1},q_{3}^{1},-q_{1}^{1}-q_{2}^{1}-q_{3}^{1},0\mathbf{)}$, $%
\mathbf{q^{2}}=\mathbf{(}%
q_{1}^{2},q_{2}^{2},q_{3}^{2},-q_{1}^{2}-q_{2}^{2}-q_{3}^{3},0\mathbf{)}$
and $\mathbf{q^{3}}=\mathbf{(}%
q_{1}^{3},q_{2}^{3},q_{3}^{3},-q_{1}^{3}-q_{2}^{3}-q_{3}^{3},0\mathbf{)}$%
\textbf{. }In these relations we have set $q_{5}^{a}=0$, a useful feature
which correspond to working in \ $\mathcal{U}_{\alpha }\left[ z_{1},\dots
,z_{4},z_{5}=1\right] $,\ the local coordinate patch of $\mathcal{Q}$ where $%
z_{5}=1$. For illustrating applications, we will mainly use the following
special choice $q_{1}^{a}=1,$ $q_{2}^{1}=q_{3}^{2}=q_{4}^{3}=-1$; all
remaining ones are equal to zero. In other words;

\begin{equation}
\mathbf{q^{1}}=\mathbf{(}1,-1,0,0,0\mathbf{);\quad q^{2}}=\mathbf{(}%
1,0,-1,0,0\mathbf{);\quad q^{3}}=\mathbf{(}1,0,0,-1,0\mathbf{).}  \label{23}
\end{equation}
The $q_{i}^{a}$ charges in eqs(\ref{21}) and eqs(\ref{23}) are defined
modulo five; $q_{i}^{a}\equiv q_{i}^{a}+5\mathbf{Z}$, and satisfy naturally
the identities

\begin{equation}
\sum_{i=1}^{5}q_{i}^{a}=0\ \ \text{modulo(5)};\ a=1,2,3.  \label{24}
\end{equation}
This constraint eq is a necessary condition required by invariance under eqs(%
\ref{21}) of the $a_{0}\prod_{i=1}^{5}z_{i}$\ monomials of eq(\ref{20}). It
ensures that the holomorphic hypersurface (\ref{20}) is indeed a CY
manifold.\ Note in passing that the constraint eq(\ref{24}) is just the
analogue of the vanishing of the first Chern class of the quintic ($c_{1}(%
\mathcal{Q})=0$)\ in differential geometry language. It is also the
condition under which the underlying 2d effective field theory flows in the
infrared to a CFT$_{2}$ \cite{34,35}. Orbifolds of the quintic with respect
to $Z_{5}^{3}$\ are as usual obtained by identifying points that are related
under the transformations (\ref{21}). Here we will show that such orbifolds
are completely characterized by the $q_{i}^{a}$\ charges and an SL$\left(
3,Z\right) $ matrix $\eta _{ab}$ and for general hypersurfaces by matrices
in SL$\left( n,Z\right) $. For symmetric matrices $\eta _{ab}$; that is \ $%
\eta _{ab}=\eta _{ba}$, one gets orbifolds without discrete torsion while
for non symmetric $\eta _{ab}$s, there is a discrete torsion. The idea
behind this classification is that together with eqs(\ref{21}-\ref{24}),
there exist extra symmetries of eqs(\ref{20}) acting as $z_{i}\rightarrow
z_{i}\omega ^{p_{i}^{a}}$ where now the $p_{i}^{a}$\ charges are given by,
\begin{equation}
p_{i}^{a}=\eta _{ab}\text{ }q_{i}^{b},  \label{25}
\end{equation}
with $\eta _{ab}$\ the above mentioned $3\times 3$\ matrix of \ $SL\left( 3;%
\mathbf{Z}\right) $. These dual charges satisfy $\sum_{i=1}^{5}p_{i}^{a}=0$,
following naturally from eq(\ref{24}). The $\eta _{ab}$ matrix, which to our
knowledge was not known before; turns out to play an important role in
building NC geometries \`{a} la BL. It appears here as encoding matrix of
the automorphisms of characters of the orbifold group. We will show later
that $\eta _{ab}$\ is the carrier of the discrete torsion of the orbifold
symmetry of CY Hypersurfaces. It antisymmetric part $\left( \eta _{ab}-\eta
_{ba}\right) $ is related to the logarithm of the $t_{ab}$\ torsion matrix
of $\mathbf{Z}_{n+2}^{n}$. Moreover, as far as $\mathcal{Q}$ is concerned,
it interesting to note that there are different kinds of orbifolds one can
build depending on the orbifold group. If we denote by $\mathcal{R}_{5}\left[
[z_{1},\dots ,z_{5}\right] ]$; $\mathcal{R}_{5}$ for short, the ring of
complex holomorphic and homogeneous polynomials of degree five on $\mathbf{CP%
}^{4}$ and by $G_{\left[ \nu \right] }$, a generic subgroup of $\mathbf{Z}%
_{5}^{3}$, then one can build various orbifolds of the quintic as $\mathcal{Q%
}^{\left[ \nu \right] }=\mathcal{R}_{5}/G_{\left[ \nu \right] }$. In
addition to $\mathcal{R}_{5}/\mathbf{Z}_{5}^{3}$, we have also the two
following examples of the quintic orbifolds $\mathcal{Q}^{\left[ 1\right] }$
and $\mathcal{Q}^{\left[ 2\right] }$ associated respectively with\ $G_{\left[
1\right] }=\mathbf{Z}_{5}$\ and $G_{\left[ 2\right] }=\mathbf{Z}_{5}^{2}$\
subgroups of $G=\mathbf{Z}_{5}^{3}$,
\begin{equation}
\mathcal{Q}^{\left[ 1\right] }=\mathcal{R}_{5}/\mathbf{Z}_{5},\qquad
\mathcal{Q}^{\left[ 2\right] }=\mathcal{R}_{5}/\mathbf{Z}_{5}^{2}.
\label{27}
\end{equation}
Throughout this study, we will mainly stay in the coordinate patch\ $%
\mathcal{U}_{(\alpha )}\left[ z_{1},\dots ,z_{4},z_{5}=1\right] $; the move
to an other patch of the quintic, say $\mathcal{U}_{(\beta )}\left[
w_{1},\dots ,w_{4},w_{5}=1\right] $\ \ with CY charges $r_{i}^{a}$, is
ensured by holomorphic transition functions $\phi _{(\alpha ,\beta
)}^{(r_{i}^{a},q_{i}^{a})}$ carrying appropriate $\mathbf{Z}_{5}^{3}$
charges. Note that on the coordinate patch \ $\mathcal{U}_{(\alpha )}\left[
z_{1},\dots ,z_{4},z_{5}=1\right] $, the full $\mathbf{Z}_{5}^{3}$\ orbifold
symmetry has no fix point; the only stable one, namely $(0,0,0,0,1)$, does
not belong to $\mathcal{R}_{5}/\mathbf{Z}_{5}^{3}$.\ However thinking of the
quintic orbifold$\ \mathcal{R}_{5}/\mathbf{Z}_{5}^{3}$ as either a $\mathbf{Z%
}_{5}$ orbifold of \ $\mathcal{R}_{5}/\mathbf{Z}_{5}^{2}$, that is $\
\mathcal{R}_{5}/\mathbf{Z}_{5}^{3}\sim \left( \mathcal{R}_{5}/\mathbf{Z}%
_{5}^{2}\right) /\mathbf{Z}_{5}\sim \mathcal{R}_{5}$'$/\mathbf{Z}_{5}\ $or
again as a \ $\mathbf{Z}_{5}^{2}$ \ orbifold of $\mathcal{R}_{5}/\mathbf{Z}%
_{5}$; i.e $\mathcal{R}_{5}/\mathbf{Z}_{5}^{3}\sim \left( \mathcal{R}_{5}/%
\mathbf{Z}_{5}\right) /\mathbf{Z}_{5}^{2}\sim \mathcal{R}_{5}$''$/\mathbf{Z}%
_{5}$,\ one can consider the fixed points of the respective singular spaces
of $\mathcal{R^{\prime }}_{5}/\mathbf{Z}_{5}$\ and \ $\mathcal{R^{\prime
\prime }}_{5}/\mathbf{Z}_{5}^{2}$.\ This procedure is also equivalent to set
to zero some of the CY vector charges associated with $\mathbf{Z}_{5}^{3}$
symmetry. For example, the orbifold $\mathcal{R}$''$_{5}/\mathbf{Z}_{5}$ may
also be linked to $\mathcal{R}_{5}/\mathbf{Z}_{5}$\ just by setting $\mathbf{%
q^{2}}=\mathbf{q^{3}}=\mathbf{0}$.\ In section 5, we shall use both the $%
\mathcal{R^{\prime }}_{5}/\mathbf{Z}_{5}$\ and\ $\mathcal{R^{\prime \prime }}%
_{5}/\mathbf{Z}_{5}^{2}$ spaces to study fractional branes.

\subsection{NC Quintic}

A way to get NC extensions of $\mathcal{Q}$ by using discrete torsion is to
start from the complex homogeneous hypersurface (\ref{20}) and choose the
coordinate patch $z_{5}=1$ and $q_{5}^{a}=0$. Then associate to the $\left\{
z_{1},z_{2},z_{3},z_{4}\right\} $\ local variables, the set of $5\times 5$\
matrix operators \ $\left\{ Z_{1},Z_{2},Z_{3},Z_{4}\right\} $\ and \ $Z_{5}$%
\ with the identity matrix\ $I_{id}$. The NC quintic $\mathcal{Q}^{nc}$,
associated to eqs(\ref{20}) and eqs(\ref{25}), is a NC algebra generated by
the $Z_{i}$'s. This\ is a special subalgebra of the ring of functions on the
space of matrices $\mathbb{M}at(5,\mathbf{C)}$; it reads in term of the $%
Z_{i}$ matrix generators as;
\begin{equation}
Z_{i}Z_{j}=\beta _{ij}\ Z_{j}Z_{i},\qquad Z_{i}Z_{5}=Z_{5}Z_{i},\qquad
i,j=1,...,4.  \label{28}
\end{equation}
In these eqs, $\beta _{ij}$ is an invertible matrix constrained as,
\begin{equation}
\beta _{ji}=\beta _{ij}^{-1},\quad \beta _{i1}^{5}=\beta _{i2}^{5}=\beta
_{i3}^{5}=\beta _{i4}^{5}=\beta _{i1}\beta _{i2}\beta _{i3}\beta
_{i4}=1,\quad \forall i.  \label{29}
\end{equation}
These constraint eqs reflect just the property that the commutative $%
\mathcal{Q}$ should be in the centre $\mathcal{Z}(\mathcal{Q}^{nc})$ of eqs(%
\ref{28}). In other words $\mathcal{Q}\equiv \mathcal{Z(Q}^{nc})$ or
equivalently;
\begin{equation}
\left[ Z_{j},Z_{i}^{5}\right] =0,\qquad \left[ Z_{j},\prod_{i=1}^{4}Z_{i}%
\right] =0.  \label{30}
\end{equation}
A class of solutions of eqs(\ref{29}) is obtained as follows: First
parameterize $\beta _{ij}$ as $\beta _{ij}=\omega ^{L_{ij}}$ with $\omega
=\exp i\frac{2\pi }{5}$ and $L_{ij}$\ is a $5\times 5$ antisymmetric matrix
satisfying $\sum_{i=1}^{5}L_{ij}=0$. This constraint relation comes from
invariance of the term $a_{0}z_{1}z_{2}z_{3}z_{4}z_{5}$ exactly as for the
CY condition eq(\ref{24}). This means that a possible solution for $L_{ij}$
is as,
\begin{equation}
L_{ij}=\nu _{ab}\left( q_{i}^{a}q_{j}^{b}-q_{j}^{a}q_{i}^{b}\right)
\label{34}
\end{equation}
where $\nu _{ab}$ is an arbitrary $3\times 3$ matrix of integer entries. The
matrix $L_{ij}$ is built as bi-linears of the $q_{k}^{c}$ charge vectors
ensuring automatically $\sum_{i=1}^{5}L_{ij}=0$. This is why we shall still
refer to the constraint eq $\sum_{i=1}^{5}L_{ij}=0$\ as the CY condition.
Moreover, since eq(\ref{34}) can also be written as $L_{ij}=m_{ab}\
q_{i}^{a}q_{j}^{b}$, with $m_{ab}=\nu _{ab}-\nu _{ba}$, one suspects that $%
m_{ab}$ matrix should be linked to the structure constants of the underlying
$\mathbf{Z}_{5}^{3}$ orbifold symmetry. We will show in section 6, that $%
L_{ij}$ should read in fact as;
\begin{equation}
L_{ij}=\ p_{i}^{a}q_{j}^{a}-p_{j}^{a}q_{i}^{a},  \label{35}
\end{equation}
where $p_{i}^{a}$ are as in eqs(\ref{25}). These integers are the charges of
a hidden invariance induced by discrete torsion. By analogy with the
geometric symmetry $\mathbf{Z}_{5}{}^{3}$, this symmetry can be thought of
as acting on the $z_{i}$ variables as $z_{i}\rightarrow z_{i}\omega
^{p_{i}^{a}}$, with $p_{i}^{a}=\nu _{ab}q_{i}^{b}.$ To have an idea on how
the formulas we have derived above work in practice, let us go the local
patch $z_{5}=1$ with $q_{i}^{a}$ charges as in eq(\ref{23}) and perform some
explicit calculations. Setting $m_{12}=k_{1}$, $m_{23}=k_{2}$ and $%
m_{13}=k_{3}$\ are integers modulo $5$, the $L_{ij}$ matrix reads as:

\begin{equation}
L_{ij}=\left(
\begin{array}{ccccc}
0 & k_{1}-k_{3} & -k_{1}+k_{2} & k_{3}-k_{2} & 0 \\
-k_{1}+k_{3} & 0 & k_{1} & -k_{3} & 0 \\
k_{1}-k_{2} & -k_{1} & 0 & k_{2} & 0 \\
-k_{3}+k_{2} & k_{3} & -k_{2} & 0 & 0 \\
0 & 0 & 0 & 0 & 0
\end{array}
\right) .  \label{37}
\end{equation}
In this case, the generators of the non commutative quintic $\mathcal{Q}%
^{nc} $ satisfy the following algebra:

\begin{eqnarray}
Z_{1}Z_{2} &=&\omega ^{k_{1}-k_{3}}Z_{2}Z_{1},\quad Z_{1}Z_{3}=\omega
^{-k_{1}+k_{2}}Z_{3}Z_{1},\quad Z_{1}Z_{4}=\omega ^{k_{3}-k_{2}}Z_{4}Z_{1},
\notag \\
Z_{2}Z_{3} &=&\omega ^{k_{1}}Z_{3}Z_{2},\quad Z_{2}Z_{4}=\omega
^{-k_{3}}Z_{4}Z_{2},\quad Z_{3}Z_{4}=\omega ^{k_{2}}Z_{4}Z_{3}.  \label{38}
\end{eqnarray}
Putting $\alpha =\omega ^{k_{1}-k_{3}},$ $\beta =\omega ^{k_{1}-k_{2}}$ and $%
\gamma =\omega ^{k_{3}}$, one discovers the algebra of $\cite{2}$ given by
eqs(\ref{20}). Moreover the solution of these eqs read, up to a
normalization factor, as:

\begin{eqnarray}
Z_{1} &=&x_{1}\mathbf{P}_{\omega ^{k_{1}+k_{2}+k_{3}}}\mathbf{Q}^{3},\qquad
Z_{2}=x_{2}\mathbf{P}_{\varpi ^{k_{1}}}\mathbf{Q}^{-1}  \notag \\
Z_{3} &=&x_{3}\mathbf{P}_{\varpi ^{k_{2}}}\mathbf{Q}^{-1},\qquad Z_{4}=x_{4}%
\mathbf{P}_{\varpi ^{k_{3}}}\mathbf{Q}^{-1},  \label{39}
\end{eqnarray}
where $x_{i}$\ are as in eqs(\ref{20}), $\varpi $ is the complex conjugate
of $\omega $ and where

\begin{equation}
\mathbf{P}_{\alpha }=diag(1,\alpha ,\alpha ^{2},\alpha ^{3},\alpha
^{4}),\quad \mathbf{Q}=\left(
\begin{array}{ccccc}
0 & 0 & 0 & 0 & 1 \\
1 & 0 & 0 & 0 & 0 \\
0 & 1 & 0 & 0 & 0 \\
0 & 0 & 1 & 0 & 0 \\
0 & 0 & 0 & 1 & 0
\end{array}
\right) ,  \label{40}
\end{equation}
with $\alpha \ $standing for $\omega ^{k_{1}},\omega ^{k_{2}}$,\ $\omega
^{k_{3}}$ and their products. This solution shows clearly that $Z_{i}^{5}$,
the product $\prod_{i=1}^{4}Z_{i}$\ and their linear combination are all of
them in the centre $\mathcal{Z}{(Q}^{nc})$ of the NC algebra $\mathcal{Q}%
^{nc}$.

\section{ NC Geometry and Discrete Torsion}

In quantum physics, non commutativity appears in different ways and has
various origins and different interpretations \cite{36,11,12}; see also \cite
{37,38,39}. In effective field theoretical models at very low energies, such
as in the Chern Simons model of the fractional quantum Hall effect \cite
{48,49}, NC geometry is generated by a strong constant external magnetic
field $B$ which couple the two position vectors $x^{i}\left( t\right) $ of
electrons as $B\varepsilon _{ij}x^{i}\left( t\right) \frac{\partial
x^{j}\left( t\right) }{\partial t}$. Heisenberg quantization rule leads to a
non vanishing commutator for these position vectors; i.e $\left[ x^{i},x^{j}%
\right] =i\frac{\epsilon ^{ij}}{B\nu _{L}}$, where $\nu _{L}=\frac{1}{k}$ is
the Laughlin filling factor. For large value of $B$ ($\sim 15$ Tesla), the
quantum properties of the system of electrons are described by a NC
Chern-Simons gauge theory on a D2 brane. Such a two dimensional condensed
matter phase has received recently an important interest due to similarities
with solitons built up with systems of D branes of type IIA string theory.
At very high energies, say around the Planck scale as in string theory, non
commutativity is generated by the NS $B_{\mu \nu }$ antisymmetric field and
is linked to the existence of open strings ending on $D$ branes\ with a
dynamics governed by a boundary conformal invariance \cite{12}. This issue
has been subject to much interest during the few last years in connection
with the derivation of non commutative ADHM solitons \cite{16} and the study
of the tachyon condensation by following the GMS method \cite{27,28}. In M
theory, NC geometry comes as a non trivial solution in the study of the
matrix model torus compactification. Here also NC geometry is generated by
an antisymmetric field; the eleven dimensional gauge three form field $%
C_{\mu \nu \rho }$\cite{11}. Elements $g_{1}$ and $g_{1}$ of the group of
automorphism symmetries of the matrix model on a two torus $\mathcal{T}^{2}$
are in general governed by the central relation $%
g_{1}g_{2}g_{1}^{-1}g_{2}^{-1}$ taken to be proportional to the identity
operator;\ that is $g_{1}g_{2}g_{1}^{-1}g_{2}^{-1}=\lambda $ $I_{id}$ with $%
\lambda \in \mathbf{C}^{\ast }$ as required by the Schur lemma \cite{11,14}.

In quantum mathematics, NC geometry is viewed as an algebraic structure $%
\mathcal{M}_{\hbar }\left[ X_{1},\dots ,X_{N}\right] $ going beyond the
usual $\mathcal{C}\left[ x_{1},\dots ,x_{N}\right] $ commutative one with
the ideal $\left\{ x_{i}x_{j}=x_{j}x_{i}\right\} $. Formally, the generic
commutation relations of the generators of the quantum algebra $\mathcal{M}%
_{\hbar }\left[ X_{1},\dots ,X_{N}\right] $, may be written as $X_{I}\ast
X_{J}=\mathrm{r}_{IJ}^{KL}(X)\ \ X_{K}\ast X_{L}+\ \ \mathrm{b}_{IJ}(x),$%
where $\mathrm{r}_{IJ}^{KL}(X)$ and $\mathrm{b}_{IJ}(X)$ are some
polynomials in $X_{I}$ which may be thought of as \cite{40,41};
\begin{equation}
\mathrm{r}_{IJ}^{KL}(X)\ =\delta _{I}^{L}\delta _{J}^{K}+\hbar \ \mathrm{r}%
_{IJ}^{\prime KL}+...;\quad \mathrm{b}_{IJ}(X)\ =\hbar \left( \Omega _{IJ}+%
\mathrm{f}_{IJ}^{K}\ X_{K}+...\right)
\end{equation}
In the limit $\hbar \ \rightarrow 0$, \ \ $\mathrm{r}_{IJ}^{KL}(X)\
\rightarrow \delta _{I}^{L}\delta _{J}^{K}\ $and$\ \mathrm{b}_{IJ}(X)\
\rightarrow 0$; one recovers the usual commutative structure of $\mathcal{C}%
\left[ x_{1},\dots ,x_{N}\right] .$ \ For the general cases, such for
instance\ $(\mathrm{r}_{IJ}^{KL}(x),\mathrm{b}_{IJ}(x))$\ equals to $%
(0,B_{IJ})$ or\ $(0,\mathrm{f}_{IJ}^{K}\ X_{K})$ or again $(\mathrm{R}%
_{IJ}^{KL}\ ,0)$, one gets respectively the canonical commutator, the Lie
algebra bracket and the quantum Yang-Baxter spaces \cite{36}. The NC
structure we are dealing with here corresponds to an other special situation
where $\mathrm{b}_{IJ}(x)=0$\ and,
\begin{equation}
\mathrm{r}_{IJ}^{KL}(x)=\mathrm{\beta }_{KL}\ \delta _{I}^{L}\ \delta
_{J}^{K};,
\end{equation}
with\ $\mathrm{\beta }_{IJ}$ is a root of unity. This NC geometry is
generated by the discrete torsion matrix of the orbifold group $\mathbf{Z}%
_{5}^{3}$ and has much to do with NC Fuzzy tori representations. Since
discrete torsion is involved in string compactifications on orbifolds and
twisted string sectors, one expects that this NC structure plays some role
in string theory on orbifolds and more generally in supersymmetric field
theories on orbifolds. As we shall show by explicit analysis, see section 4,
NC geometry induced by discrete torsion leads to fractional $D$ branes at
orbifold singularities and offers a new way to resolve non geometric
singularities. Points of the usual geometry are replaced by polygons in the
NC case.

To get the right link between discrete torsion and NC orbifold solutions, in
particular quintic ones given\ above, let us reconsider the solutions eqs(%
\ref{39})\ and explore their structure. To do so we shall first show that
the NC solution (\ref{39}) are not so general as claimed in $\cite{24}$ as
these solutions form just a special class of a more general set of solutions
involving general representations of $\mathbf{Z}_{5}^{3}$. We will \ then
begin by giving some results regarding regular representations of $\mathbf{Z}%
_{5}^{3}$; after that we present our general discrete torsion dependent
solution.

\subsection{More on Solutions (\ref{39})}

Eqs(\ref{39}) involve various group elements of the representation of $%
\mathbf{Z}_{5}$ namely the $\mathbf{P}_{\omega ^{k_{1}}},\mathbf{P}_{\omega
^{k_{2}}}$ and $\mathbf{P}_{\omega ^{k_{3}}}$ commuting operators and powers
of $\mathbf{Q}$. These five dimensional matrices system $\left\{ \mathbf{P}%
_{\omega ^{k_{1}}},\mathbf{P}_{\omega ^{k_{2}}},\mathbf{P}_{\omega ^{k_{3}}},%
\mathbf{Q}\right\} $ have the following torsion matrices,

\begin{equation}
\mathbf{t}_{\mu \nu }=\left(
\begin{array}{cccc}
1 & 1 & 1 & \omega ^{k_{1}} \\
1 & 1 & 1 & \omega ^{k_{2}} \\
1 & 1 & 1 & \omega ^{k_{a}} \\
\omega ^{-k_{1}} & \omega ^{-k_{2}} & \omega ^{-k_{3}} & 1
\end{array}
\right) ,\quad \mathbf{\theta }_{\mu \nu }=\left(
\begin{array}{cccc}
0 & 0 & 0 & k_{1} \\
0 & 0 & 0 & k_{2} \\
0 & 0 & 0 & k_{3} \\
-k_{1} & -k_{2} & -k_{3} & 0
\end{array}
\right)
\end{equation}
where we have set $\theta _{\mu \nu }=\frac{-5i}{2\pi }\log \mathbf{t}_{\mu
\nu }$. Therefore discrete torsion exist whenever at least one of the $k_{a}$
integers is non zero. Moreover since $\mathbf{P}_{\alpha }\mathbf{P}_{\beta
}=\mathbf{P}_{\alpha \beta }$, it follows that the expression of the $Z_{i}$%
s may also be written as $Z_{i}=x_{i}\otimes \mathbf{T}_{i},$where the $%
T_{i} $'s are five dimensional matrices realized as,
\begin{equation}
\mathbf{T}_{i}=\mathbf{P}^{r_{i}}\mathbf{Q}^{s_{i}}=\prod_{a=1}^{3}\mathbf{P}%
^{r_{i}^{a}}\mathbf{Q}^{s_{i}^{a}},\quad \mathbf{T}_{i}\mathbf{T}_{j}=%
\mathbf{\tau }_{ij}\mathbf{T}_{j}\mathbf{T}_{i},\quad \mathbf{\tau }%
_{ij}=\prod_{a=1}^{3}\omega
_{a}^{s_{i}^{a}r_{j}^{a}-s_{j}^{a}r_{i}^{a}}=\omega ^{s_{i}r_{j}-s_{j}r_{i}},
\label{037}
\end{equation}
where we have set\ $\mathbf{P=P}_{\omega }$ and\ where the numbers $%
r_{i}^{a} $ and $s_{i}^{a}$\ are integers modulo five related to the $k_{i}$
as $r_{1}=\sum_{a=1}^{3}r_{1}^{a}=k_{1}+k_{2}+k_{3}$, $s_{1}=%
\sum_{a=1}^{3}s_{1}^{a}=3$ and $r_{i}=\sum_{a=1}^{3}r_{i}^{a}=-k_{i}$, $%
s_{i}=\sum_{a=1}^{3}s_{i}^{a}=-1$ for $i=2,3,4$. One of the remarkable
features of the solution (\ref{39}) together with eq(\ref{037}) is that it
has only a manifest $\mathbf{Z}_{5}$ subsymmetry and so constitutes a
special class of realization of the NC quintic. More general solutions
should have a full manifest $\mathbf{Z}_{5}^{3}$\ symmetry. Larger class of
solutions corresponds to take the $T_{i}$ operators of eq(\ref{037}) as $%
T_{i}=\prod_{a=1}^{3}\mathbf{E}_{a}^{r_{i}^{a}}\mathbf{J}_{a}^{s_{i}^{a}}$
with $\mathbf{E}_{a}$s\ and $\mathbf{J}_{a}$s given by
\begin{eqnarray}
E_{1} &=&P_{1}\otimes I_{id}\otimes I_{id},\quad E_{2}=I_{id}\otimes
P_{2}\otimes I_{id},\quad E_{3}=I_{id}\otimes I_{id}\otimes P_{3},  \notag \\
J_{1} &=&Q_{1}\otimes I_{id}\otimes I_{id},\quad J_{2}=I_{id}\otimes
Q_{2}\otimes I_{id},\quad J_{3}=I_{id}\otimes I_{id}\otimes Q_{3}.
\label{nc0}
\end{eqnarray}
These are respectively the generators of $\mathbf{Z}_{5}^{3}$ and the group
of automorphisms of their characters. In what follows, we explore further
this general solution and give its geometric interpretation in terms of
D-branes wrapping the compact manifold.

\subsubsection{Torsion and NC rational torus}

The solution $Z_{i}=x_{i}\otimes \mathbf{T}_{i}$ may be given a remarkable
geometric interpretation. The $Z_{i}$s\ are the local coordinates of a NC
fiber bundle whose base is a Chart of $\mathbf{CP}^{4}$ and its fiber is a
NC rational torus $\mathcal{T}^{2}$. Generic points of this NC variety are
then parameterized as $\left( x_{i};P_{\omega _{1}},Q_{\omega
_{1}};P_{\omega _{2}},Q_{\omega _{2}};P_{\omega _{3}},Q_{\omega _{3}}\right)
$\ where now $P_{\omega _{a}}$ and $Q_{\omega _{a}}$ are viewed as the
cycles of a rational torus
\begin{equation}
\mathbf{P}_{\omega _{a}}^{5}=\mathbf{Q}_{\omega _{a}}^{5}=I,\qquad \mathbf{Q}%
_{\omega _{a}}\mathbf{P}_{\omega _{a}}=\omega _{a}\ \mathbf{P}_{\omega _{a}}%
\mathbf{Q}_{\omega _{a}}  \label{nct}
\end{equation}
and where $\omega _{a}^{5}=1$. This representation shows clearly that
discrete torsion\ associated with each $\mathbf{Z}_{5}$\ subgroup factor
induces a 2d NC torus fibration of the quintic. Since this 2d NC rational\
torus is finite dimensional\footnote{%
Note that for the special limit of irrational tori corresponding to taking
the $k\rightarrow \infty $ limit of $Z_{k}\sim U\left( 1\right) $, non
compact extra dimensions appear. The infinite number of fractional\ $Dp$ is
mapped to a non compact $Dp+2$ brane in agreement with the result of \cite
{46,47}.}, the original commutative coordinates $z_{i}1$ are now replaced by
$5\times 5$ matrices as
\begin{equation}
z_{i}1=\ \rightarrow \ \ Z_{i}=\sum_{k,l=1}^{5}\ Z_{i}^{kl}\ |k><l|,
\label{nctt}
\end{equation}
where we have used the $\left\{ e_{ij}=|i><j|\right\} $ matrix basis$\left\{
|i>\right\} $ and $\left\{ <j|\right\} $, with $<j|i>=\delta _{ij}$. In the
case the three $\mathbf{Z}_{5}$ factors are taken into account, one has a 6d
NC torus fibration. In this case, the \ matrices are of order $5^{3}$ and
then the k and l indices in the expansion (\ref{nctt}) should be thought of
as multi-indices; that is $k=\left( k_{1},k_{2},k_{3}\right) $\ and $%
l=\left( l_{1},l_{2},l_{3}\right) $.

\subsubsection{Branes}

Due to discrete torsion, we see form the above eq that the algebraic
structure of the D-branes wrapping the compact manifold change. Brane points
$\{z_{i}\}$ of commutative geometry become now fibers based on $\{z_{i}\}$
and valued in the\ group representation $\mathcal{D}\left( Z_{5}^{3}\right) $
as shown on eq(\ref{nctt}). Following \cite{17}, this solution has a nice
interpretation in terms of quiver diagrams. Associating to each \ $\mathbf{e}%
_{kl}$\ matrix vector basis, a segment $\left[ k,l\right] $ oriented from $k$
to $l$ (an arrow $\overrightarrow{kl}$) and to each $\mathbf{e}_{kk}={\pi
_{k}}\equiv |k><k|$ projector, a loop starting and ending at the position $k$
as shown on the following table, one may draw a quiver diagram for each $%
Z_{i}$ matrix generator of the NC algebra.

\begin{equation}
\begin{tabular}{|l|l|}
\hline
Operators & \ \ \ \ Diagrams \\ \hline
$\mathbf{a}_{k}^{+}\equiv |k><k+1|$ & $\ \ \ \ _{k}\quad \bullet $ $%
\longrightarrow \bullet \quad _{\left( k+1\right) }$ \\ \hline
$\mathbf{a}_{k}^{-}\equiv |k+1><k|$ & $\ \ \ _{\left( k+1\right) }\quad
\bullet $ $\longleftarrow \bullet \quad _{k}$ \\ \hline
$\prod_{j=0}^{n}\mathbf{a}_{k+j}^{+}\equiv |k><k+n|$ & $\ \ \ \ _{k}\quad
\bullet \longrightarrow \bullet \quad _{\left( k+n\right) }$ \\ \hline
$\prod_{j=0}^{n}\mathbf{a}_{k+j}^{-}\equiv |k+n><k|$ & $\ \ \ _{\left(
k+n\right) }\quad \bullet \longleftarrow \bullet \quad _{k}$ \\ \hline
$\pi _{k}=\mathbf{a}_{k}^{+}\mathbf{a}_{k}^{-}=|k><k|$ & $\ \ \ \
_{k}\bullet $ $\leftrightharpoons $ $\bullet $ $\ _{\left( k+1\right) }\quad
\equiv \ _{k}\bigcirc \quad \equiv \bullet $ \\ \hline
$\pi _{k}=\mathbf{a}_{k-1}^{-}\mathbf{a}_{k-1}^{+}=|k><k|$ & $\ \ \ \
_{\left( k-1\right) }\bullet $ $\leftrightharpoons $ $\bullet $ $\ _{k}\quad
\equiv \ \bigcirc $ $_{k}\quad \equiv \bullet $ \\ \hline
\end{tabular}
\end{equation}
Using these rules, one sees that for the quintic the generic quiver diagram
for the $Z_{i}$\ operators is given by a polygon with five vertices (a
pentagon) and in general twenty links joining the various vertices; see
figure 1. Since matrix basis vectors type $\mathbf{e}_{k\left( k+n\right)
}=|k><k+n|$ can be usually decomposed as $\mathbf{e}_{k\left( k+n\right) }=%
\mathbf{a}_{k}^{+}\mathbf{a}_{k+1}^{+}...\mathbf{a}_{k+n}^{+},$where $%
\mathbf{a}_{k}^{+}=\mathbf{e}_{k\left( k+1\right) }$, one concludes that
points in NC quintic geometry are roughly speaking described by pentagons.
\begin{figure}[tbh]
\begin{center}
\epsfxsize=6cm \epsffile{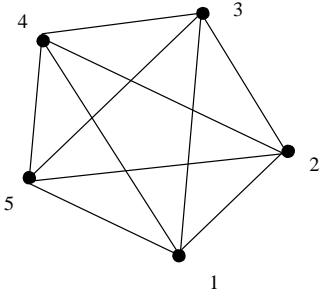}
\end{center}
\par
\caption{{\protect\small \textit{This diagram represents a point in the NC
quintic $\mathcal{Q}^{nc}$. As generic links may usually be decomposed in
terms of the $|j><j+1|$ boundary edges, this quiver diagram can be thought
as a pentagon; see also fig 2.}}}
\end{figure}
On the NC quintic the D-branes acquire an internal structure and
consequently singularities of the original commutative manifolds are
resolved by NC geometry.

\subsection{Discrete Torsion Matrix of $\mathbf{Z}_{5}^{3}$}

Here we want to study general $\mathcal{D}\left( \mathbf{Z}_{5}^{3}\right) $
representations in presence of torsion and derive their link with complex
three dimension NC rational torus $\mathcal{T}_{\omega }^{6}$. We also give
the quiver diagrams associated with these representations. To that purpose,
we shall first consider the simplest situation where all $\mathbf{Z}_{5}$
factors commute amongst themselves; then we discuss the case where they do
not commute.

\subsubsection{Free torsion case}

Naively, the geometric symmetry $\mathbf{Z}_{5}^{3}$ can be seen as the
product of three abelian $\mathbf{Z}_{5}$ group factors whose generators may
be defined by help of the tensor product as follows:

\begin{equation}
E_{1}=P_{1}\otimes I_{id}\otimes I_{id},\quad E_{2}=I_{id}\otimes
P_{2}\otimes I_{id},\quad E_{3}=I_{id}\otimes I_{id}\otimes P_{3},
\label{038}
\end{equation}
The $E_{a}$s are the generators of the three\ $\mathbf{Z}_{5}$\ factors of \
$\mathbf{Z}_{5}^{3}$; they satisfy the cyclic property $E_{a}^{5}=\mathbf{I}%
_{\mathcal{D}\left( G\right) }$ \ following from the individual identities $%
P_{a}^{5}=I_{id}$. Since they are commuting operators, $%
E_{a}.E_{b}=E_{b}.E_{a}$, they can be diagonalized simultaneously in the
same basis \ $\{|a,i>;1\leq a\leq 3;\ 1\leq i\leq 5\}$. As such the $P_{a}$s
can be thought of as in eq(\ref{40}) and $E_{a}$s as diagonal blocks of
matrices. Using the convention notation $\mathbf{a}_{n_{a}}^{+}=|a,n><a,n+1|$%
,$\ \mathbf{a}_{n_{a}}^{-}=\ |a,n+1><a,n|$, $\ \pi _{n_{a}}\ =\ \mathbf{a}%
_{n_{a}}^{+}\ \mathbf{a}_{n_{a}}^{-}$ and the graphic representations of
figure 2, it is not difficult to see that group elements of $\mathbf{Z}_{5}$%
\ and their automorphisms can be decomposed as follows,

\begin{equation}
I_{id}=\sum_{n=1}^{5}\pi _{n},\quad P_{a}=\sum_{n=1}^{5}\alpha _{a,n}\pi
_{n},\quad Q_{a}=\sum_{n=1}^{5}\mathbf{a}_{a,n}^{+};\quad
Q_{a}^{-1}=\sum_{n=1}^{5}\mathbf{a}_{a,n}^{-},
\end{equation}
For explicit computations, we shall drop out the index $a$ by working in
special five dimensional matrix realizations.

\begin{figure}[tbh]
\begin{center}
\epsfxsize=7cm \epsffile{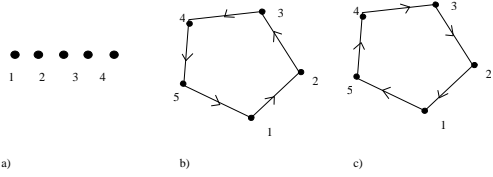}
\end{center}
\caption{{\protect\small \textit{The completely reducible diagram (Fig2a)
represents the identity operator of 5d representation $\mathcal{D}(\mathbf{Z}%
_{5})$. Fig2b is an oriented pentagon representing $Q$ automorphism operator
while fig2c \ is its inverse. }}}
\end{figure}
Note that in absence of torsion generic elements of\ $\mathbf{Z}_{5}^{3}$
are denoted as $g=g_{1}\otimes g_{2}\otimes g_{3}$ and similarly for their
representations $\mathcal{D}\left( g\right) =\mathcal{D}\left( g_{1}\right)
\otimes \mathcal{D}\left( g_{2}\right) \otimes \mathcal{D}\left(
g_{3}\right) $ which read\ in terms of the $E_{a}$ generators as follows $%
\mathcal{D}\left( g\right) =\prod_{i_{1},i_{2},i_{3}=1}^{5}\ \gamma
_{i_{1}i_{2}i_{3}}E_{1}^{i_{1}}E_{2}^{i_{2}}E_{3}^{i_{3}}$, where the $%
\gamma _{i_{1}i_{2}i_{3}}$ coefficients are such that\ $\ {\gamma
_{i_{1}i_{2}i_{3}}}^{5}=1$. Since the group multiplication law $gg^{\prime }$%
\ of elements $g$ and $g^{\prime }$\ of $\mathbf{Z}_{5}^{3}$\ is defined as
usual by performing multiplications of individual elements; that is $%
gg^{\prime }=g_{1}g_{1}^{\prime }\otimes g_{2}g_{2}^{\prime }\otimes
g_{3}g_{3}^{\prime }$; we have $E_{a}^{5}=\mathbf{I}_{\mathcal{D}\left(
\mathbf{Z}_{5}^{3}\right) }$, with $\mathbf{I}_{\mathcal{D}\left( \mathbf{Z}%
_{5}^{3}\right) }$ stands for $I_{id}\otimes I_{id}\otimes I_{id}$; the\
group representation identity. The above eq tells us that the dimension $d$
of $\mathcal{D}\left( \mathbf{Z}_{5}^{3}\right) $ reads in terms of the $%
d_{i}$ dimensions of the three $\mathcal{D}\left( \mathbf{Z}_{5}\right) $
factors as $d=d_{1}d_{2}d_{3}$. As all elements of\ $\mathbf{Z}_{5}^{3}$\
can be expressed as powers of ${E_{a}}$s, we will focus our attention now on
the monomials $E_{{1}}^{n_{1}}E_{2}^{n_{2}}E_{3}^{n_{3}}$; $1\leq n_{a}\leq
5 $.

\subsubsection{ Discrete Torsion}

In presence of discrete torsion\footnote{%
With one $\mathbf{Z}_{5}$ factor, one has discrete torsion induced by
quantum symmetry. Here we discuss the general case of discrete torsion
between the various geometric orbifold subgroup factors of \ $\mathbf{Z}%
_{5}^{3}$.}, the $g_{a}$ elements of the three $\mathbf{Z}_{5}$ factors of $%
\mathbf{Z}_{5}^{3}$ do not commute among each others. Geometrically this
situation corresponds to the case where the three complex cycles $\mathcal{T}%
_{\omega _{a}}^{2}$ associated with discrete group factors do not commute.
Generic couples $\left( F_{a},F_{b}\right) $\ of elements of $\mathcal{T}%
_{\omega _{a}}^{2}\otimes \mathcal{T}_{\omega _{b}}^{2}$ satisfy then,
\begin{equation}
F_{a}F_{b}=t_{ab}F_{b}F_{a},  \label{tor}
\end{equation}
where $t_{ab}$\ is the torsion matrix between the three $\mathbf{Z}_{5}$
factors of the orbifold group. Note that this relation is quite similar to
the one defining the rational torus $\mathcal{T}_{\omega _{1}}^{2}$ eq(\ref
{nct}). Instead of one $\mathbf{Z}_{5}$, we have now the full $\mathbf{Z}%
_{5}^{3}$ orbifold group. Eq(\ref{tor})define a complex three dimension NC
rational torus $\mathcal{T}_{\vartheta }^{6}$ where a priori the six real
cycles are non commuting.\ Moreover, like for the $E_{a}$s, the $F_{a}$s
satisfy equally $F_{a}^{5}=\mathbf{I}_{\mathcal{D}\left( \mathbf{Z}%
_{5}^{3}\right) }$ requiring that the matrix torsion should be of the form $%
t_{ab}=\exp i\frac{2\pi }{5}\theta _{ab},$where $\theta _{ab}$ is
antisymmetric $3\times 3$ matrix with integer entries.

\textbf{Representations}\textsl{\ \qquad }There are different ways to
represent eq(\ref{tor}) by using tensor products or/and direct sums
involving the $d_{a}$ dimension matrix generators $E_{a}$ of the three $%
\mathcal{D}\left( \mathbf{Z}_{5}\right) $ and the $Q_{a}$ automorphisms
rotating their characters. In the case where one uses tensor products, the
matrix representation we get has dimension $d=d_{1}d_{2}d_{3}$ containing as
a particular case the solutions obtained in \cite{2,24}. To see how this
representation is built, we introduce the following parameterization of
torsion $\theta _{ab}=\eta _{ab}-\eta _{ba}$where $\eta _{ab}$ is the
invertible $SL(3,\mathbf{Z})$ matrix considered before. Note in passing that
the fact that $\eta _{ab}$ belongs to $SL(3,\mathbf{Z})$ appears here as a
necessary condition for consistency; but this may have a string
interpretation in terms of allowed values of the NS B field along the
rational elliptic fibers. This parameterization allows us to rewrite eq(\ref
{tor}) as $\omega ^{\eta _{ba}}F_{a}F_{b}=\omega ^{\eta _{ab}}F_{b}F_{a}$
whose matrix solution reads as
\begin{equation}
F_{1}=P_{1}\otimes Q_{2}^{\eta _{12}}\otimes Q_{3}^{\eta _{13}},\quad
F_{2}=Q_{1}^{\eta _{21}}\otimes P_{2}\otimes Q_{3}^{\eta _{23}},\quad
F_{3}=Q_{1}^{\eta _{31}}\otimes Q_{2}^{\eta _{32}}\otimes P_{3},
\label{expf}
\end{equation}
where the $P_{a}$s and the $Q_{a}$s are as in eq(\ref{40}) with \ $%
P_{a}Q_{b}=\alpha _{a}\ Q_{b}P_{a}\delta _{ab}$ and $\alpha _{a}$ fifth
roots of unity. Note that for\ symmetric $\eta _{ab}$, the $F_{a}$s commute
as seen on (\ref{sym}) and the orbifold has no discrete torsion. To fix the
ideas, we set for simplicity $\alpha _{1}=\alpha _{2}=\alpha _{3}=\omega $
and consider the special case where the three basis of the three
representation factors of $\mathcal{D}\left( \mathbf{Z}_{5}\right) $, namely
$\{|a,i>\ ;1\leq i\leq d_{a};\ \,1\leq a\leq 3\}$, have the same dimension $%
d_{1}=d_{2}=d_{3}$. In this case $d_{1}=d_{2}=d_{3}=5$ and so $%
P_{1}=P_{2}=P_{3}=P$ and $Q_{1}=Q_{2}=Q_{3}=Q$ with a realization as in eqs(%
\ref{nct}). Thus the $F_{a}$s\ reduce to:
\begin{equation}
F_{1}=P\otimes Q^{\eta _{12}}\otimes Q^{\eta _{13}},\quad F_{2}=Q^{\eta
_{21}}\otimes P\otimes Q^{\eta _{23}},\quad F_{3}=Q^{\eta _{31}}\otimes
Q^{\eta _{32}}\otimes P.
\end{equation}
This matrix representation has an a $5^{3}$ order and satisfies $F_{a}^{5}=I$%
. It extends eqs(\ref{038}) which appears as special cases. Indeed, the $d$
dimensional generic representations of the $F_{a}$s shows that it is
possible to build different realizations for $F_{a}$ generators (\ref{expf}%
). The representation eqs(\ref{38}) and (\ref{037}) built in $\cite{24}$,
correspond to take $(d_{1},d_{2},d_{3})$ equal to either $(5,1,1)$,$\
(1,5,1) $, or $(1,1,5)$ respectively obtained by setting $\eta _{a2}=\eta
_{a3}=0,\ \eta _{a1}=\eta _{a3}=0$ \ and $\eta _{a1}=\eta _{a2}=0$. In all
of these cases, the fiber of the NC quintic is just the NC rational torus $%
\mathcal{T}_{\omega _{a}}^{2}$. This property may be viewed as the geometric
interpretation of the codimension one fixed planes considered in \cite{17}.
In addition to these examples, there are other special cases such as\ the $%
25 $ dimensional matrix realizations of the $F_{a}$s. They correspond to the
situations where the fibration is $\mathcal{T}_{\theta }^{4}$ and $%
(d_{1},d_{2},d_{3})=(5,5,1)$ as well as permutations.

\textbf{Quiver Diagrams}\textsl{\ \qquad }Like for the case of one abelian
factor, one can also build the projectors for full the \ $\mathbf{Z}_{5}^{3}$
\ group. Using the individual $\mathbf{Z}_{5}$ projectors $\pi _{k_{a}}={%
\frac{1}{5}}\sum_{i=1}^{5}\ {\omega ^{-k_{a}}}\ \mathbf{P}_{\omega }^{i}$,
we can construct various kinds of projectors on the representation space of $%
\mathbf{Z}_{5}^{3}$. First, the${\ \Pi }_{k_{a}}$ projectors on the $\mathbf{%
Z}_{5}$ representation spaces;
\begin{equation}
{\Pi }_{k_{1}}\ =\ {\pi _{k_{1}}}\otimes I_{id}\otimes I_{id},\quad {\ \Pi }%
_{k_{2}}\ =\ I_{id}\otimes \pi _{k_{2}}\otimes I_{id},\quad {\ \Pi }%
_{k_{3}}\ =\ I_{id}\otimes I_{id}\otimes \pi _{k_{3}},
\end{equation}
They have quiver diagrams more a less similar to that of ${\pi _{k_{a}}}$,
except that now we have general realizations coming from the decomposition
of the identity operators. The full quiver diagram is given by the cross
product of the individual graphs and one ends with higher dimensional
lattices. Second, the${\ \Pi }_{(k_{a},k_{b})}$ and${\ \Pi }%
_{(k_{1},k_{2},k_{3})}$ projectors on the $\mathbf{Z}_{5}^{2}$ \ and $%
\mathbf{Z}_{5}^{3}$ representation spaces, respectively obtained by taking
tensor products of ${\ \Pi }_{k_{a}}^{\prime }s$;
\begin{equation}
{\Pi }_{(k_{a},k_{b})}\ ={\Pi }_{k_{a}}{\Pi }_{k_{b}},\quad {\Pi }%
_{(k_{1},k_{2},k_{3})}={\Pi }_{k_{1}}{\Pi }_{k_{2}}{\Pi }_{k_{3}}.
\end{equation}
Accordingly, the identity matrix $\mathbf{I}_{\mathcal{D}\left( G\right) }$
can be decomposed in different, but equivalent, ways as shown here below;
\begin{equation}
\mathbf{I}_{\mathcal{D}\left( G\right) }=\sum_{k_{a}=1}^{5}{\Pi }%
_{k_{a}}=\sum_{k_{a},k_{b}=1}^{5}{\Pi }_{k_{a},k_{b}}=%
\sum_{k_{1},k_{2},k_{3}=1}^{5}{\Pi }_{k_{1}}{\Pi }_{k_{2}}{\Pi }_{k_{3}}.
\label{040}
\end{equation}
So $\mathbf{I}_{\mathcal{D}}$ can be represented by a completely reducible
quiver diagram with $d_{1}d_{2}d_{3}=5\times 5\times 5$ vertices. Similar
expansion to eqs(\ref{040}) may be written down for the generators $J_{a}$
of the quantum symmetries. Setting
\begin{equation}
A_{k_{1}}^{\pm }=a_{1,k_{1}}^{\pm }\otimes I_{id}\otimes I_{id},\quad
A_{k_{2}}^{\pm }=I_{id}\otimes a_{1,k_{1}}^{\pm }\otimes I_{id},\quad
A_{k_{3}}^{\pm }=I_{id}\otimes I_{id}\otimes a_{k_{3}}^{\pm },
\end{equation}
we can write for instance $J_{a}$\ and \ $J_{a_{1}}J_{a_{2}}$ as follows;
\begin{equation}
J_{a}=\sum_{k_{a}=1}^{5}\ A_{k_{a}}^{+},\quad
J_{a_{1}}J_{a_{2}}=\sum_{k_{a_{1}}k_{a_{2}}=1}^{5}\
A_{k_{a_{1}}}^{+}A_{k_{a_{2}}}^{+}.
\end{equation}
While the quiver diagram for the $J_{a}$\ is similar to that given by
figures 2b, 2c and figures 3, the quiver diagrams associated with\ $%
J_{a_{i}}J_{a_{j}}$\ are obtained by taking cross products and are of type
figure 4.
\begin{figure}[tbh]
\begin{center}
\epsfxsize=12cm \epsffile{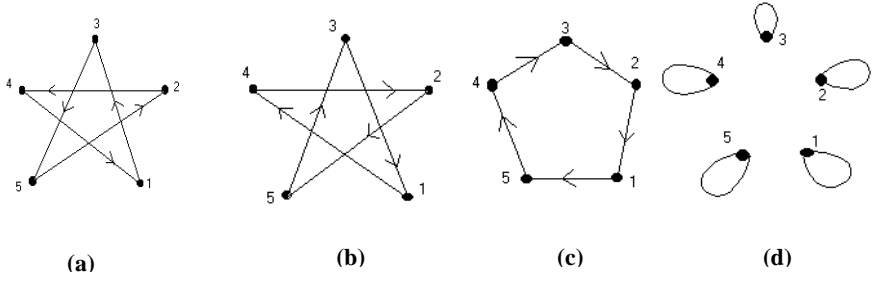}
\end{center}
\caption{{\protect\small \textit{Fig3a represents the diagram of $Q^{2}$;
the oriented links define the independent massless chiral fields on the $D$
brane at singularity. Fig3b represents the quiver of $Q^{3}$. Fig 3c
represents the diagram of the $Q^{4}$ operator and Fig3d represents the
completely reducible quiver diagram of $Q^{5}$. }}}
\end{figure}

\section{General Solutions}

Here we give our general solutions for NCCY orbifolds extending the ones
obtained in \cite{2,24}. These solutions exhibit manifestly both torsion
dependence and the full orbifold geometric symmetry. The solution we will
derive provide novel regularizations of NC field theories on orbifolds
containing as special models gauge theories embedded in string theory. To
avoid repetitions, we will treat simultaneously the example quintic
threefolds $\mathcal{Q}$ and more generally all the elements of the class of
homogeneous complex $n$ ($n>1$) dimension hypersurfaces $\mathcal{H}_{n}$.
To start recall that the algebraic relations defining the NC quintic $%
\mathcal{Q}^{(nc)}$ as appeared first in \cite{2} reads as,

\begin{eqnarray}
Z_{1}Z_{2} &=&\alpha Z_{2}Z_{1},\quad Z_{1}Z_{3}=\alpha ^{-1}\beta
Z_{3}Z_{1},\quad Z_{1}Z_{4}=\beta ^{-1}Z_{4}Z_{1},\quad Z_{2}Z_{3}=\alpha
\gamma Z_{3}Z_{2},  \notag \\
Z_{2}Z_{4} &=&\gamma ^{-1}Z_{4}Z_{2},\quad Z_{3}Z_{4}=\beta \gamma
Z_{4}Z_{3},\quad Z_{i}Z_{5}=Z_{5}Z_{i},\qquad i=1,2,3,4;  \label{qint}
\end{eqnarray}
where $\alpha ,\beta ,\gamma $ are fifth roots of unity. In \cite{24}, it
was noted that the above relations are very special and can generalized to
any complex $n$-dimension holomorphic homogeneous CY hypersurfaces as $%
Z_{i}Z_{j}=\beta _{ij}Z_{j}Z_{i}$, $\ i,j=1,...,(n+1)$ and $%
Z_{i}Z_{d+2}=Z_{d+2}Z_{i}\ $with $i=1,...,(n+1)$ and $\beta _{kl}$s are
realized as,

\begin{equation}
\beta _{ij}=\exp i\left( \frac{2\pi }{n+2}m_{ab}q_{i}^{a}q_{j}^{b}\right)
=\omega ^{m_{ab}q_{i}^{a}q_{j}^{b}}.  \label{betaj}
\end{equation}
In this relation $\omega =\exp i\frac{2\pi }{n+2}$ and$\ m_{ab}$ is some
given matrix with integer entries which remained without interpretation in
\cite{24}. Here, we will prove that $m_{ab}$ is equal to
\begin{equation}
m_{ab}=\eta _{ab}{}^{-1}-\eta _{ba}{}^{-1}  \label{ma}
\end{equation}
where $\eta _{ab}$\ is as before. The antisymmetric part of the $\eta _{ab}$
matrix encodes then torsion and it is required to belong to SL$\left(
n,Z\right) $.

\subsection{More on NC Quintic}

First of all note that the NC quintic and more generally NCCY hypersurfaces
are no uniquely defined. The following eqs may be also used as definitions
for $\mathcal{Q}^{nc}$ and $\mathcal{H}_{n}^{\left( nc\right) }$,
\begin{eqnarray}
\Phi _{i}\Phi _{j} &=&\Phi _{j}\Phi _{i},\quad i,j=1,...,n+2  \notag \\
F_{a}\Phi _{i} &=&\Phi _{i}F_{a}\exp i\left( \frac{2\pi }{n+2}%
q_{i}^{a}\right) ,  \label{qint2} \\
F_{a}F_{b} &=&F_{b}F_{a}\exp i\left( \frac{2\pi }{n+2}\theta _{ab}\right) .
\notag
\end{eqnarray}
These eqs have also a centre that coincide exactly with $\mathcal{H}_{n}$.
Therefore the two sets of matrix coordinates $Z_{i}$ and $\Phi _{i}$, of eqs(%
\ref{qint},\ref{betaj}) and (\ref{qint2}) should be linked. In fact as shown
in \cite{17}, $Z_{i}$ and $\Phi _{i}$ are two Morita equivalent coordinates
of $\mathcal{H}_{n}^{\left( nc\right) }$ and so are related as
\begin{equation}
Z_{i}=\Gamma _{i}\Phi _{i},  \label{auto}
\end{equation}
where $\Gamma _{i}$s are matrix operators which can be directly derived by
comparing eqs(\ref{betaj}) and (\ref{qint2}). We will give their explicit
expressions later on. For the moment, let us comment the absence of a
relation such that $F_{a}\Phi _{i}=\Phi _{i}F_{a}\exp i\frac{2\pi q_{i}^{a}}{%
5}$ in eqs(\ref{qint},\ref{betaj}). At first sight this seems a little bit
ambiguous as the naive counting of the degrees of freedom in relations (\ref
{betaj}) and (\ref{qint2}) do not match. However, this is not a problem
since though the $F_{a}$ group generators do not appear manifestly in eqs (%
\ref{qint},\ref{betaj}); they act as outer automorphisms on these eqs. It
turns out that $F_{a}$ act trivially on $Z_{i}$; that is $%
F_{a}Z_{i}F_{a}^{-1}=Z_{i}$ \ which can be also rewritten like,
\begin{equation}
F_{a}Z_{i}=Z_{i}F_{a}  \label{constr}
\end{equation}
This relation will play an important role in building the general explicit
fiber dependent solutions of eqs(\ref{betaj},\ref{qint2}). Indeed as $%
F_{a}\Phi _{i}=\Phi _{i}F_{a}\omega ^{q_{i}^{a}}$ is just the Morita
transformation of (\ref{constr}), one can use it to determine $\Gamma _{i}$.
Acting by $F_{a}$ on (\ref{constr}) and using (\ref{auto}), one gets the
following constraint on $\Gamma _{i}$; $F_{a}\Gamma _{i}=\omega ^{-\nu
_{ia}}\Gamma _{i}F_{a},$where $\nu _{ia}$ are integers.

\textbf{Solutions of NC algebra }$Z_{i}Z_{j}=\beta _{ij}Z_{j}Z_{i}$:\textsl{%
\qquad }Using the matrix realization of the $E_{a}$ and $J_{a}$ generators
of the rational torus fibers eqs(\ref{nc0}) and (\ref{40}) as well as the
relations $E_{a}^{p_{i}^{a}}J_{b}^{q_{j}^{b}}=\delta _{ab}\omega
^{p_{i}^{a}q_{j}^{b}}J_{b}^{q_{j}^{b}}E_{a}^{p_{i}^{a}}$, it is not
difficult to check that the $Z_{i}$s eq(\ref{betaj})\ are solved as,
\begin{equation}
Z_{i}=\prod_{a=1}^{n}E_{a}^{p_{i}^{a}}J_{a}^{q_{i}^{a}}.  \label{salz}
\end{equation}
Computing the products $Z_{i}Z_{j}$ and $Z_{j}Z_{i}$, one gets the explicit
expression of the $\beta _{ij}$\ parameters namely,
\begin{equation}
\beta _{ij}=\exp i\frac{2\pi }{n+2}\sum_{a=1}^{n}\
[p_{i}^{a}q_{j}^{a}-p_{j}^{a}q_{i}^{a}].  \label{betak}
\end{equation}
Next, solving the constraint eq $F_{a}Z_{i}=Z_{i}F_{a}$ by using eq(\ref
{salz}) and the explicit expression of the $F_{a}$s namely $%
F_{a}=\prod_{b=1}^{n}E_{a}J_{b}^{\eta _{ab}}$, we find:

\begin{equation}
\sum_{b=1}^{n}\eta _{ab}p_{i}^{b}=q_{i}^{a}.  \label{ps}
\end{equation}
This relation shows that the $p_{i}^{a}$s are related to the $q_{i}^{a}$s
via the torsion matrix as $p_{i}^{a}=\sum_{b=1}^{3}\eta _{ab}^{-1}q_{i}^{b}$%
. Since the $p_{i}^{a}$s\ are integers, eq(\ref{ps}) requires that the
matrix $\eta _{ab}$ has to belong to$\ SL(n,\mathbf{Z})$ and shows moreover
that $p_{i}^{a}$s satisfy themselves the identity,

\begin{equation}
\sum_{i=1}^{n}p_{i}^{a}=0.  \label{yp}
\end{equation}
This eqs(\ref{yp}) tells us whenever torsion is present, the orbifold of the
hypersurface $\mathcal{H}_{n}$ admits an extra hidden discrete symmetry
acting on the $z_{i}$'s as $z_{i}\rightarrow z_{i}\exp i\frac{2\pi }{n+2}%
\eta _{ab}^{-1}q_{i}^{b}$. This eq requires $\eta _{ab}$\ to be invertible
and can be viewed a geometric way to define orbifolds. For $\eta _{ab}^{-1}$
antisymmetric, the orbifolds have then a discrete torsion. Finally using eq(%
\ref{ps}) and comparing eq(\ref{betak}) with eq(\ref{betaj}), one discovers
eq(\ref{ma}).

\textbf{Solution of eqs}(\ref{qint2}):\textsl{\qquad }Since the $\Phi _{i}$s
commute, a natural solution corresponds is to take $\Phi _{i}$ as depending
uniquely of the $E_{a}$s or again uniquely of $J_{a}$ generators. For the
second case for instance, the\ $\Phi _{i}$s are realized as,
\begin{equation}
\Phi _{i}=z_{i}\prod_{a=1}^{n}J_{a}^{q_{i}^{a}}.  \label{salf}
\end{equation}
In addition to commutativity, this representation fulfills naturally $%
F_{a}\Phi _{i}=\Phi _{i}F_{a}\exp i\frac{2\pi q_{i}^{a}}{n+2}$ due to the
basic relation $E_{a}^{p_{i}^{a}}J_{b}^{q_{j}^{b}}=\delta _{ab}\omega
^{p_{i}^{a}q_{j}^{b}}J_{b}^{q_{j}^{b}}E_{a}^{p_{i}^{a}}$. Moreover,
comparing the two representations (\ref{salz}) and (\ref{salf}), one gets
the expression of the automorphisms $\Gamma _{i}$ of eq(\ref{auto}),
\begin{equation}
\Gamma _{i}=\prod_{a=1}^{n}E_{a}^{p_{i}^{a}},  \label{aut}
\end{equation}
where the $p_{i}^{a}$s are as in eqs(\ref{ps}).

\textbf{More on Morita equivalence:}\textsl{\qquad }We end this discussion
by noting that given a set of $q_{i}^{a}$ integers, defining the charges of
the $z_{i}$ variables under the ${\mathbf{Z}_{n+2}^{n}}$, and non symmetric $%
n\times n$ matrix $\eta _{ab}$ of \ $SL(n;Z)$; we can build various, but
Morita equivalent, realizations of the NC algebra describing $\mathcal{H}%
_{n}^{nc}$ where eqs(\ref{salz}) and (\ref{salf}) appear as two special
coordinates basis amongst many others. If we let $\left\{
F_{a},W_{i}\right\} $ a generic basis of $\mathcal{H}_{n}^{nc}$ related to $%
\left\{ F_{a},\Phi _{i}\right\} $ as $W_{i}=\Omega _{i}$ $\Phi _{i}$
\footnote{%
More general relations use $W_{i}=\Omega _{ij}$ $\Phi _{j}$.} where $\Omega
_{i}=\Omega _{i}(E_{a},J_{a})$ are constrained as,
\begin{equation}
\Omega _{i}\Omega _{j}=\varepsilon _{ij}\ \Omega _{j}\text{ }\Omega
_{i},\quad F_{a}\Omega _{i}=\kappa _{ai}^{-1}\ \Omega _{i}\text{ }F_{a}.
\label{mor1}
\end{equation}
In these eqs, $\varepsilon _{ij}$ and $\kappa _{ai}$ are some given phases
satisfying $\varepsilon _{ij}^{n+2}=\kappa _{ai}^{n+2}=1$. In this basis,
the defining relations of the NC algebra for $\mathcal{H}_{n}^{nc}$ reads as,

\begin{equation}
W_{i}W_{j}=\tau _{ij}^{-1}\tau _{ij}W_{j}W_{i},\quad F_{a}W_{i}=\kappa
_{ai}^{-1}\lambda _{ai}W_{i}F_{a},\quad F_{a}F_{b}=F_{b}F_{a}\exp i\frac{%
2\pi \theta _{ab}}{n+2},  \label{mor2}
\end{equation}
where $\tau _{ij}$, $\kappa _{ai}$ and $\lambda _{ai}$\ are structure
constants of the $\mathcal{H}_{n}^{nc}$. In these eqs, the $W_{i}$
generators do no longer commute among themselves nor with the group
representation generators $F_{a}$ as it was the case for eqs(\ref{salz}) and
(\ref{salf}) which are recovered as two extreme situations.

\subsection{Example}

To fix the ideas, let consider the example of the quintic by\ choosing\ $%
\Omega _{i}$\ as$\ F_{1}^{r_{1}^{a}}F_{2}^{r_{2}^{a}}F_{3}^{r_{3}^{a}}$\
with $r_{i}^{a}$ some given integers. In this case,\ the structure constants
appearing in eqs(\ref{mor1}) and (\ref{mor2}) read as;
\begin{equation}
\tau _{i}=\ \omega ^{\sum_{a}{r_{i}^{a}q_{i}^{a}}},\quad \varepsilon _{ij}=\
\omega ^{\theta _{ab\text{ }}r_{i}^{a}\text{ }r_{j}^{b}},\quad \kappa
_{ai}=\ \omega ^{-\theta _{ab}\ {r_{i}^{b}}}.
\end{equation}
From these relations we see that for $\eta _{ab}\ $symmetric, the structure
constants $\varepsilon _{ij}$ and $\kappa _{ai}$ are torsion free\ ($%
\varepsilon _{ij}=1$\ and $\kappa _{ai}=1$). Note that in the $%
\{W_{i};F_{a}\}$ basis, the generators of the NC algebra of the quintic do
not commute in general; except for the two following special cases where
they take remarkable forms: (a) $\{F_{a};\Phi _{i}\}$\textbf{\ }basis which
is recovered by choosing the structure constants as $\tau
_{i}^{-1}\varepsilon _{ij}\tau _{j}=1$; i.e,
\begin{equation}
\varepsilon _{ij}=\tau _{i}\tau _{j}^{-1},\quad \kappa _{ai}=1.
\end{equation}
In this case, the matrix $\Omega _{i}$\ is just the inverse of\ $\Gamma _{i}$%
; that is $\Omega _{i}=\Gamma _{i}^{-1}$ eq(\ref{aut}). (b) $\{F_{a};Z_{i}\}$%
\textbf{\ }basis obtained from the generic\ $\left\{ F_{a},W_{i}\right\} $
frame by requiring commuting {\textbf{$Z_{i}$}}\ and $F_{a}$\ operators.
This is equivalent to seting\ $\kappa _{ai}\sigma _{ai}=1$ which imply in
turns,

\begin{equation}
\sigma _{ai}=\kappa _{ai}^{-1},\quad \beta _{ij}=\tau _{i}^{-1}\varepsilon
_{ij}\tau _{j},\quad \Omega _{i}=\Gamma _{i}.
\end{equation}
Since the two sets of matrix generators\ $\{Z_{i}\}$ and $\{F_{a}\}$
decouple completely, the NC quintic $\mathcal{A}[Z_{i};F_{a}]$ is then
described by a trivial fibration as $\mathcal{A}[Z_{i};F_{a}]\equiv \mathcal{%
A}[Z_{i}]\otimes \mathcal{A}[F_{a}]$.

\section{Fractional Branes}

The realization of the NC quintic we have studied here above concerns only
the regular points of the algebra, that is non singular ones. In this
section, we want to complete this analysis by considering the
representations for singular points. This is not only important for the
study of fractional branes at singularities\ but also for answering the
question regarding the nature of fractional branes on the NC quintic and
more generally on NCCY hypersurfaces. To do so, we shall first determine the
various sets $\mathcal{S}_{(\mu )}$ of singular points of orbifolds of the
quintic. Then we give the corresponding singular solutions. At first sight
and as far as the full $\mathbf{Z}_{5}^{3}$ geometric symmetry is concerned
we have only one fixed point under the $\mathbf{Z}_{5}^{3}$ actions namely $%
(z_{1},z_{2},z_{3},z_{4},1)=(0,0,0,0,1).$ This point belongs however to the $%
\mathbf{CP}^{4}$ projective space; but does not belong to the quintic $%
\mathcal{Q}$; no point of the quintic is then fixed by the full symmetry.
This property is valid for all CY hypersurfaces; no point of complex $n$
dimensional CY hypersurfaces $\mathcal{P}{(z_{1},\dots ,z_{n+2})}$\ is fixed
under the full $\mathbf{Z}_{n+2}^{n}$ geometric invariance. We will
therefore consider points of $\mathcal{Q}$ that are fixed under subgroups $%
G_{\left[ \alpha \right] }$ of $\ \mathbf{Z}_{5}^{3}$. Then we describe the
various fractional branes living at these singularities, the corresponding
quiver diagrams and the massless chiral fields of the effective theory on
the D branes. As there are several subgroups $G_{\left[ \alpha \right] }$ in
$\mathbf{Z}_{5}^{3}$, we shall fix our attention on two categories of
subsymmetries; those isomorphic to\ $\mathbf{Z}_{5}$;\ i.e $G_{\left[ 1%
\right] }\simeq \mathbf{Z}_{5}$ and those isomorphic to $\mathbf{Z}_{5}^{2}$%
;\ i.e\ $G_{\left[ 2\right] }\simeq \mathbf{Z}_{5}^{2}$. The CY charges will
be taken as in eqs(\ref{23}). Generalization to subgroups of $\mathbf{Z}%
_{n+2}^{n}$, though tedious, is a priori straightforward.

\subsection{Fixed subspaces of $\mathcal{Q}^{\left[ \protect\alpha \right] }$%
}

We will consider first the spaces $\mathcal{S}_{(a)}$ of \ fixed points
under a generic $\mathbf{Z}_{5}$ factor of $\mathbf{Z}_{5}^{3}$. Then we
examine the spaces $\mathcal{S}_{(ab)}$ of fixed points under$\ \mathbf{Z}%
_{5}^{2}$ factors. To have an idea on how these spaces look like, it is
interesting to think about the quintic homogeneous hypersurface eq(\ref{20})
as a fiber bundle described by the following equation $P(z_{1},\ldots
,z_{5})=\sum_{n_{m+1}...n_{m+5}=0}^{5}\ b_{n_{m+1}...n_{m+5}}\
z_{i_{m+1}}^{n_{m+1}}\dots z_{i_{m+5}}^{n_{m+5}}$, where $%
b_{n_{m+1}...n_{m+5}}=b_{n_{m+1}...n_{m+5}}(z_{i_{1}},\ldots ,z_{i_{m}})$
are some given monomials in the $z_{i_{1}},\ldots ,z_{i_{m}}$ complex
variables, with appropriate degrees. Let us give examples on how this works
in practice.

\subsubsection{$\mathbf{CP}^{2}\bowtie \mathcal{S}_{1}$\textit{\ }Fibration}

A simple example for realizing fibrations of the quintic consists to rewrite
$P(z_{1},\ldots ,z_{5})$ as,

\begin{equation}
P(z_{1},\ldots ,z_{5})=b_{00}+b_{11}\text{ }z_{1}z_{2}+b_{50}\text{ }%
z_{1}^{5}+b_{05}\text{ }z_{2}^{5},  \label{043}
\end{equation}
where the $b_{mn}$\ coefficients are holomorphic functions given by
\begin{eqnarray}
b_{00}(z_{3},z_{4},z_{5}) &=&z_{3}^{5}+z_{4}^{5}+z_{5}^{5},\quad
b_{11}(z_{3},z_{4},z_{5})=\ a_{0}z_{3}.z_{4}.z_{5},  \notag \\
b_{50}(z_{3},z_{4},z_{5}) &=&1,\quad b_{05}(z_{3},z_{4},z_{5})=1.
\label{044}
\end{eqnarray}
The remaining others are equal to zero. Eqs(\ref{043},\ref{044}) mean that
the quintic may be viewed as a fibration space with base $\mathbf{CP}^{2}$
and fiber $\mathcal{S}_{1}$ given by,

\begin{equation}
z_{1}^{5}+z_{2}^{5}+b_{1}\ z_{1}z_{2}=0.  \label{045}
\end{equation}
This relation is invariant under the change\ $\left( \ z_{1},\text{ }%
z_{2}\right) \rightarrow \left( \ \omega \text{ }z_{1},\text{ }\omega
^{-1}z_{2}\right) $; that is under the $\mathbf{Z}_{5}$\ subsymmetry of
charges $q_{i}^{1}$; the $\mathbf{B}_{1}$ base is not affected under this
change. The symmetry of the\ $\mathcal{S}_{1}$\ fiber has one fixed point
namely $\left( 0,0\right) $ and so\ $\mathcal{S}_{1}$ is singular at the
origin$\ z_{1}=z_{2}=0$. To see what eq(\ref{045}) represents, note that
from the $z_{1}$\ and $z_{2}$\ variables, one can build three invariant
namely $u=z_{1}^{5}$, $v=z_{2}^{5}$ and $\ w=z_{1}z_{2}$ having an $\mathbf{A%
}_{4}$\ singularity. In terms of the new variables, the equation of the $%
\mathcal{S}_{1}$\ complex curve reads as
\begin{equation}
u+v+b\text{ }w=0;\quad uv=w^{5}.  \label{046}
\end{equation}
Therefore near the fixed point $z_{1}=z_{2}=0$, the $\mathbf{Z}_{5}$\
orbifold of the commutative quintic $\mathcal{Q}$ can be then viewed as
given by the fiber bundle $\mathbf{CP}^{2}\bowtie \mathcal{S}_{1}$ with a
vanishing two cycle at $z_{1}=z_{2}=0$. Before going ahead, let us comment
briefly the complex resolution of this kind of singularity and give its
toric geometry diagram representation as shown here below,

\begin{equation}
\begin{tabular}{|l|l|}
\hline
$\mathbf{A}_{4}$ Singularity & $uv=w^{5}$ \\ \hline
\begin{tabular}{l}
Complex \\
Resolution of $\mathbf{A}_{4}$%
\end{tabular}
& $uv=w^{5}+\alpha _{4}w^{4}+\alpha _{3}w^{3}+\alpha _{2}w^{2}+\alpha
_{1}w+\alpha _{0}$ \\ \hline
Rules &
\begin{tabular}{l}
$i)$ \ \ White\ nodes such as\ $\ \bigcirc $ \ , are associated \\
\ \ \ \ \ to each non compact $\ \mathbf{C}$ variables $x$ and $y$ \\
$ii)$ \ \ Nodes such as \ $\otimes $ \ are associated with \\
\ \ \ \ \ \ blown up spheres with self intersection $\left( -2\right) $ \\
$iii)$ \ \ Each link\ \ \ $\longleftrightarrow $\ \ represents intersecting
\\
\ \ \ \ \ \ \ spheres with a weight $\left( 1\right) $%
\end{tabular}
\\ \hline
Quiver Diagrams &
\begin{tabular}{l}
$i)$ \ Quiver diagram for the resolution of $\mathbf{A}_{2}$: \\
\ \ \ \ \ \ \ \ \ \ \ \ \ \ $\bigcirc _{z^{2}}\longrightarrow \otimes
_{z}\longleftarrow \bigcirc _{1}$ \\
$ii)$ \ Quiver diagram for the resolution of $\mathbf{A}_{4}$ \\
$\bigcirc _{w^{5}}\longleftrightarrow \otimes _{w^{4}}\longleftrightarrow
\otimes _{w^{3}}\longleftrightarrow \otimes _{w^{2}}\longleftrightarrow
\otimes _{w}\longleftarrow \bigcirc $ $_{1}$%
\end{tabular}
\\ \hline
\end{tabular}
\end{equation}
More details on this graphic representation of Kahler and complex resolution
of $ADE$ singularities as well as their applications in string
compactifications may be found in\cite{42,45}.

\subsubsection{$\mathbf{B}\bowtie \mathcal{S}_{12}$\textit{\ \ }Fibration}

The second example we want to give corresponds to the fibration $\mathbf{B}%
\bowtie \mathbf{F}\equiv {\mathbf{CP}^{1}\bowtie }\mathcal{S}_{12}$. In this
case, the analogue of the above equations read for $\mathcal{S}_{12}$ as
follows:

\begin{equation}
P(z_{1},\dots ,z_{5})=b_{000}+b_{111}\text{ }z_{1}z_{2}z_{3}+b_{500}\text{ }%
z_{1}^{5}+b_{050}\text{ }z_{2}^{5}+b_{005}\text{ }z_{3}^{5},  \label{047}
\end{equation}
where the $b_{mnr}$\ coefficients are as follows,
\begin{eqnarray}
b_{000}(z_{4},z_{5}) &=&z_{4}^{5}+z_{5}^{5},\quad b_{111}(z_{4},z_{5})=\
a_{0}z_{4}.z_{5},  \notag \\
b_{500}(z_{4},z_{5}) &=&1,\quad b_{050}(z_{4},z_{5})=1,\quad
b_{005}(z_{4},z_{5})=1,  \label{048}
\end{eqnarray}
and all others are equal to zero. The equation of the singular fiber $%
\mathcal{S}_{12}$ is given by $z_{1}^{5}+z_{2}^{5}+z_{3}^{5}+b\
z_{1}z_{2}z_{3}=0$; this is a complex surface which has one fixed fiber
point at $\left( 0,0,0\right) $ under the change $\left(
z_{1,}z_{2},z_{3}\right) \longrightarrow \left( \omega ^{2}z_{1,}\omega
^{-1}z_{2},\omega ^{-1}z_{3}\right) $ generating the $\mathbf{Z}_{5}\otimes
\mathbf{Z}_{5}$\ subsymmetry with charges $\left( q_{i}^{1},q_{i}^{2}\right)
$. Note that this subsymmetry does not affect the \ $\mathbf{B}_{12}$ base
space. Now introducing the following four invariant $u_{1}=z_{1}^{5}$, $%
u_{2}=z_{2}^{5}$, $u_{3}=z_{3}^{5}$ and $t=z_{1}z_{2}z_{3}$, one sees that
they are related as
\begin{equation}
u_{1}u_{2}u_{3}=t^{5};  \label{101}
\end{equation}
while the $\mathcal{S}_{12}$ complex surface reads in terms of these
invariants as $u_{1}+u_{2}+u_{3}+bt=0$. From this relation, one recognizes
the two individual singularities associated with each factor of the $\mathbf{%
Z}_{5}\otimes \mathbf{Z}_{5}$ symmetry. These are given by the following $%
\mathbf{A}_{4}$\ eqs;
\begin{eqnarray}
u_{1}u_{2} &=&\frac{t^{5}}{u_{3}},\quad \text{for \ }u_{3}\neq 0,  \notag \\
u_{1}u_{3} &=&\frac{t^{5}}{u_{2}},\quad \text{for \ }u_{2}\neq 0.
\end{eqnarray}
Eq(\ref{101}) describes the case where both of the above singularities
collapse; it has a nice description in terms of quiver diagrams.

\begin{equation}
\begin{tabular}{|l|l|}
\hline
Singularity & $u_{1}u_{2}u_{3}=t^{5}$ \\ \hline
\begin{tabular}{l}
Complex \\
Resolution
\end{tabular}
& $u_{1}u_{2}u_{3}=t^{5}+\alpha _{4}t^{4}+\alpha _{3}t^{3}+\alpha
_{2}t^{2}+\alpha _{1}t+\alpha _{0}$ \\ \hline
Rules & Same rules as in previous table. \\ \hline
Symmetries &
\begin{tabular}{l}
\\
$\mathbf{Z}_{5}:\qquad \ \ z_{1}\rightarrow \omega z_{1}$, $z_{3}\rightarrow
\omega ^{-1}z_{3}$,$\quad \quad z_{1}z_{3}$ \ is an invariant. \\
$\mathbf{Z}_{5}\otimes \mathbf{Z}_{5}:z_{1}\rightarrow \omega ^{2}z_{1}$, $%
z_{2}\rightarrow \omega ^{-1}z_{2}$, $z_{3}\rightarrow \omega ^{-1}z_{3}$,
\\
$\mathbf{Z}_{5}:\qquad \ \ z_{1}\rightarrow \omega z_{1}$, $z_{2}\rightarrow
\omega ^{-1}z_{2}$,\quad $\quad z_{1}z_{2}$ \ is an invariant.
\end{tabular}
\\ \hline
\begin{tabular}{l}
Quiver \\
Diagram
\end{tabular}
&
\begin{tabular}{l}
\\
$\ \ \ \ \ _{\swarrow }\otimes _{\frac{t^{5}}{z_{1}z_{2}}%
}\longleftrightarrow \otimes _{\frac{t^{5}}{\left( z_{1}z_{2}\right) ^{2}}%
}\longleftrightarrow \otimes _{\frac{t^{5}}{\left( z_{1}z_{2}\right) ^{3}}%
}\longleftrightarrow \otimes _{\frac{t^{5}}{\left( z_{1}z_{2}\right) ^{4}}%
}\leftrightarrow \bigcirc $ $_{u_{3}}$:$\mathbf{Z}_{5}$ \\
$\bigcirc _{t^{5}}\longleftrightarrow \otimes _{t^{4}}\longleftrightarrow
\otimes _{t^{3}}\longleftrightarrow \otimes _{t^{2}}\longleftrightarrow
\otimes _{t}\leftrightarrow \bigcirc _{1}$:$\qquad \qquad \mathbf{Z}_{5}^{2}$
\\
\ \ \ \ \ $^{\nwarrow }\otimes _{\frac{t^{5}}{z_{1}z_{3}}%
}\longleftrightarrow \otimes _{\frac{t^{5}}{\left( z_{1}z_{3}\right) ^{2}}%
}\longleftrightarrow \otimes _{\frac{t^{5}}{\left( z_{1}z_{3}\right) ^{3}}%
}\longleftrightarrow \otimes _{\frac{t^{5}}{\left( z_{1}z_{3}\right) ^{4}}%
}\leftrightarrow \bigcirc $ $_{u_{2}}$:$\mathbf{Z}_{5}$ \\
\\
Here it is represented the three graphs associated to the resolution \\
of the singularities of the discrete symmetries reported above.
\end{tabular}
\\ \hline
\end{tabular}
\end{equation}
Note that the $\mathbf{Z}_{5}\otimes \mathbf{Z}_{5}$\ symmetry has a total
charge charge $\mathbf{q}^{1}+\mathbf{q}^{2}=\left( 2,-1,-1,0,0\right) $,
behaving then as the $\mathbf{Z}_{5}$ diagonal symmetry. The remaining off
diagonal factor has a total charge $\mathbf{q}^{1}-\mathbf{q}^{2}=\left(
0,-1,1,0,0\right) $.

\subsubsection{Other Fibrations}

Following the same method we have used for $\mathcal{S}_{1}$ and $\mathcal{S}%
_{12}$, we can work out the other $\mathbf{B}_{a}\bowtie $\ $\mathcal{S}_{a}$
and \ $\mathbf{B}_{\left( ab\right) }\bowtie $\ $\mathcal{S}_{\left(
ab\right) }$\ quintic fibrations associated with the natural subgroups of $%
\mathbf{Z}_{5}^{3}$. Denoting the various invariants under the subgroups of $%
\mathbf{Z}_{5}^{3}$ as $u_{i}=z_{i}^{5}$, $w_{ij}=z_{i}z_{j}$ and $\
t_{ijk}=z_{i}z_{j}z_{k}$, one can work out the different equations of the $%
\mathcal{S}_{a}$ and fibers $\mathcal{S}_{\left( ab\right) }$; the basis $%
\mathbf{B}_{a}$\ and $\ \mathbf{B}_{\left( ab\right) }$ are respectively
given by the $\mathbf{CP}^{2}$ and $\mathbf{CP}^{1}$ complex projective
spaces. The results are collected in the table eq(\ref{202}),

\begin{equation}
\begin{tabular}{|l|l||l|l|}
\hline
Fibers & Equations & Fibers & Equations \\ \hline
$\mathcal{S}_{1}$ &
\begin{tabular}{l}
$u_{1}+u_{2}+b_{1}\text{ }w_{12}=0$ \\
$u_{1}u_{2}=w_{12}^{5}.$%
\end{tabular}
\qquad & $\mathcal{S}_{12}$ &
\begin{tabular}{l}
$u_{1}+u_{2}+u_{3}+b$ $t_{123}=0$, \\
$u_{1}u_{2}u_{3}=t_{123}^{5}$.
\end{tabular}
\\ \hline
$\mathcal{S}_{2}$ &
\begin{tabular}{l}
$u_{1}+u_{3}+b_{2}\text{ }w_{13}=0$ \\
$u_{1}u_{3}=w_{13}^{5}.$%
\end{tabular}
\qquad & $\mathcal{S}_{23}$ &
\begin{tabular}{l}
$u_{1}+u_{3}+u_{4}+bt_{134}=0$, \\
$u_{1}u_{3}u_{4}=t_{134}^{5}$.
\end{tabular}
\\ \hline
$\mathcal{S}_{3}$ &
\begin{tabular}{l}
$u_{1}+u_{4}+b_{3}\text{ }w_{14}=0$ \\
$u_{1}u_{4}=w_{14}^{5}.$%
\end{tabular}
\qquad & $\mathcal{S}_{13}$ &
\begin{tabular}{l}
$u_{1}+u_{2}+u_{4}+bt_{124}=0$, \\
$u_{1}u_{2}u_{4}=t_{124}^{5}$.
\end{tabular}
\\ \hline
\end{tabular}
\label{202}
\end{equation}
Having these results at hand, we turn now to give some details by studying
the fixed spaces under the orbifold subgroups $\mathbf{Z}_{5}$\ and $\mathbf{%
Z}_{5}^{2}$. We first consider the orbifolds $\mathcal{Q}^{\left[ 1\right]
}\simeq \mathcal{R^{\prime }}/\mathbf{Z}_{5}$ and then the \ $\mathbf{Z}%
_{5}^{2}$ orbifolds $\mathcal{Q}^{\left[ 2\right] }\simeq \mathcal{R^{\prime
\prime }}/\mathbf{Z}_{5}^{2}$. These orbifolds correspond also to start from
eq(\ref{20}) and choose either one of the three $q_{i}^{a}$ vector charges
non vanishing say $q_{i}^{1}=\left( 1,-1,0,0,0\right) $ while the two others
$\left( q_{i}^{2}\right) =\left( q_{i}^{3}\right) =\mathbf{0}$;\ or two
vector charges non vanishing while the third is zero such as for instance\ $%
q_{i}^{1}=\left( 1,-1,0,0,0\right) ,$\ $q_{i}^{2}=\left( 1,0,-1,0,0\right) $%
\ and\ $\left( q_{i}^{3}\right) =\mathbf{0}$.

\subsection{ Fractional Branes on $\mathcal{Q}^{\left[ 1\right] }$}

As there are three manifest $\mathbf{Z}_{5}$ subsymmetry factors in the
orbifold group $\mathbf{Z}_{5}^{3}$, each one generated by an operator $%
F_{a} $, one can write down three corresponding $\mathbf{B}_{a}\bowtie $\ $%
\mathcal{S}_{a}$ fibrations for the commutative quintic.\ The $\mathbf{B}%
_{a} $ spaces are the bases of the fibration and $\ $the $\mathcal{S}_{a}$s
their fibers.\newline
\textbf{Example }: $\mathcal{Q}^{\left[ 1\right] }\simeq \mathbf{B}%
_{1}\bowtie $\ $\mathcal{S}_{1}$

Consider the $\mathbf{Z}_{5}$\ subgroup generated by $E_{1}$ with $q_{i}^{1}$
charges taken as $q_{i}^{1}=\left( 1,-1,0,0,0\right) $; the $\mathbf{B}%
_{1}\bowtie $\ $\mathcal{S}_{1}$ fibration of the quintic is just that given
by eqs(\ref{043}-\ref{045}). Since we are working in the coordinate patch $%
z_{5}=1$, the $\mathbf{B}_{1}$\ base is a patch of $\mathbf{CP}^{2}$; that
is $\mathbf{B}_{1}\sim \mathbf{C}^{2}$ parameterized by the $z_{3}$ and $%
z_{4}$ complex coordinates. The codimension two fiber $\mathcal{S}_{1}$ is
given by the following complex\ curve with an $\mathbf{A}_{4}$ singularity,
\begin{equation}
u+v+\frac{v^{5}}{u}=0.
\end{equation}
\textbf{Results}\textsl{\ }

The $\mathbf{B}_{a}\bowtie $\ $\mathcal{S}_{a}$\ fibrations of the quintic
are completely determined by the $q_{i}^{a}$ CY charges. The\ $\mathbf{B}%
_{a} $\ base manifolds are parameterized by those holomorphic coordinates $%
z_{i}$ with $q_{i}^{a}=0$ while the $\mathcal{S}_{a}$ fibers are
parameterized by those complex variables with non zero $q_{i}^{a}$ charges
with fixed points at the origin. For the $q_{i}^{a}$ 's taken as in eq(\ref
{23}), we have the following results collected in the table eq(\ref{203}).

\begin{equation}
\begin{tabular}{|l|l|l|l|}
\hline
Generators $F_{a}$ & $\ \ \ \ \ \ \ \ \ \ \ \ F_{1}$ & $\ \ \ \ \ \ \ \
F_{2} $ & $\ \ \ \ \ \ \ \ \ \ F_{3}$ \\ \hline
Fixed Points & $\left( 0,0,z_{3},z_{4},z_{5}\right) $ & $\left(
0,z_{2},0,z_{4},z_{5}\right) $ & $\left( 0,z_{2},z_{3},0,z_{5}\right) $ \\
\hline
$\mathbf{B}_{a}=\mathbf{CP}^{2}$ & $\mathbf{B}_{12}=\left\{
z_{3},z_{4},z_{5}\right\} $ & $\mathbf{B}_{23}=\left\{
z_{2},z_{4},z_{5}\right\} $ & $\mathbf{B}_{31}=\left\{
z_{2},z_{3},z_{5}\right\} $ \\ \hline
$\mathcal{S}_{a}$ Fibers &
\begin{tabular}{l}
$u_{1}+w_{12}+\frac{w_{12}^{5}}{u_{1}}=0$ \\
$u_{1}=z_{1}^{5}$, $u_{2}=z_{2}^{5}$, \\
$w_{13}=z_{1}z_{3}$.
\end{tabular}
&
\begin{tabular}{l}
$u_{1}+w_{13}+\frac{w_{13}^{5}}{u_{1}}=0$ \\
$u_{1}=z_{1}^{5}$, $u_{3}=z_{3}^{5}$, \\
$w_{12}=z_{1}z_{2}$.
\end{tabular}
&
\begin{tabular}{l}
$u_{1}+w_{14}+\frac{w_{14}^{5}}{u_{1}}=0$ \\
$u_{1}=z_{1}^{5}$, $u_{4}=z_{4}^{5}$, \\
$w_{14}=z_{1}z_{4}$.
\end{tabular}
\\ \hline
\end{tabular}
\label{203}
\end{equation}
Having the above features in mind, the singular representations of the NC
quintic may be obtained by starting from the regular representation eqs(\ref
{salf}) and taking the appropriate limits. For the $\mathcal{S}_{1}$
singular space for instance, the field moduli are obtained from eqs(\ref
{salf}) by setting $q_{i}^{1}=q_{i}^{2}=0$ \ and taking the zero limit of $\
z_{1}$ and $z_{2}$. We have for the fiber $\mathcal{S}_{1}$,

\begin{equation}
\Phi _{1}=\ \left( {{Q}\otimes }I_{id}\otimes I_{id}\right) \
\lim_{z_{1}\rightarrow 0}\ {z_{1}},\quad \Phi _{2}=\ \left( {{Q^{-1}}}%
\otimes I_{id}\otimes I_{id}\right) \ \lim_{z_{1}\rightarrow 0}\ {z_{2}}%
,\quad F_{1}=P\otimes Q^{\eta _{12}}\otimes Q^{\eta _{13}},
\end{equation}
and for $\mathbf{B}_{1}$ base,

\begin{eqnarray}
\Phi _{3} &=&z_{3}\ \left( I_{id}{\otimes }I_{id}\otimes I_{id}\right)
,\quad \Phi _{4}=z_{4}\ \left( I_{id}{\otimes }I_{id}\otimes I_{id}\right)
,\quad \Phi _{5}=z_{5}\ \left( I_{id}{\otimes }I_{id}\otimes I_{id}\right) ,
\notag \\
F_{2} &=&Q^{\eta _{21}}\otimes P\otimes Q^{\eta _{23}},\quad F_{3}=Q^{\eta
_{31}}\otimes Q^{\eta _{32}}\otimes P.
\end{eqnarray}
As one sees, once the limit to the singular point is taken, the non
vanishing matrix field moduli $\Phi _{i}$,\ $i=3,4,5$ \ are proportional to
the identity$\ \left( I_{id}{\otimes }I_{id}\otimes I_{id}\right) =\mathbf{I}%
_{\mathcal{D}\left( G\right) }$\ of the group representation\ $\mathcal{D}%
\left( G\right) $; i.e \ $\Phi _{i}=z_{i}\otimes \mathbf{I}_{\mathcal{D}%
\left( G\right) }$. This is a very remarkable feature at singularity and has
algebraic and brane interpretations.\newline
\textbf{Fractional branes on}\textit{\ }$\mathcal{R^{\prime }}/\mathbf{Z}%
_{5} $

To fix the ideas, let start from the $D9$ brane of type $IIB$ string wrapped
on the quintic $Q$. Let $\left( x^{\mu };\text{ }y_{1}\text{, }y_{2}\text{, }%
y_{3}\right) $ denote the $D9$ coordinates with $x^{\mu }=\left( x^{0},\text{
}x^{1}\text{, }x^{2},\text{ }x^{3}\right) $ being the longitudinal non
compact variables ( representing a $D3$ brane embedded in $D9$) and the $%
y_{i}$'s the compact transverse complex coordinates of the wrapped $D9$
branes ($D9\sim D3\times \mathcal{Q}$). In the coordinate patch $%
z_{5}=z_{4}=1$, the $y$ coordinates may be imagined as related to those of
the quintic as,
\begin{equation}
y_{1}=z_{1},\quad y_{2}=z_{2},\quad y_{3}=\frac{-a_{0}z_{1}z_{2}z_{3}}{%
\left( 2+z_{1}^{5}+z_{2}^{5}+z_{3}^{5}\right) }.  \label{rcor}
\end{equation}
In the case of $Q^{\left[ 1\right] }$ with discrete torsion, the wrapped $D9$
becomes a NC brane generated by the algebra eqs(\ref{salf}). At the fixed
point of $Q^{\left[ 1\right] }$, a real two cycle shrinks to zero and the
original NC wrapped $D9$ \ reduces to a NC wrapped $D7$ described by;

\begin{eqnarray}
\Phi _{3} &=&z_{3}\ {I}_{\mathcal{D}\left( \mathbf{Z}_{5}\right) },\qquad
\Phi _{4}=z_{4}\ {I}_{\mathcal{D}\left( \mathbf{Z}_{5}\right) },\qquad \Phi
_{5}=z_{5}\ {I}_{\mathcal{D}\left( \mathbf{Z}_{5}\right) },  \notag \\
F_{1} &=&P\text{,}\qquad F_{2}=Q^{\eta _{21}},\qquad F_{3}=Q^{\eta _{31}}%
\text{.}  \label{q1}
\end{eqnarray}
The singular modes at the orbifold point are carried by the ${Q}$ operator
and its inverse ${{Q^{-1}}}$ as shown on the following eqs.

\begin{equation}
\Phi _{1}=\ {Q}\text{ }\ \lim_{z_{1}\rightarrow 0}\ {z_{1}},\qquad \Phi
_{2}=\ {{Q^{-1}}}\ \lim_{z_{2}\rightarrow 0}\ {z_{2}}.
\end{equation}
However due to the complete reducibility property of ${I}_{\mathcal{D}\left(
\mathbf{Z}_{5}\right) }$ namely ${I}_{\mathcal{D}\left( \mathbf{Z}%
_{5}\right) }=\sum_{n=1}^{5}\Pi _{n}$,\ eqs(\ref{q1}) describe in fact a set
of five commuting Euclidean wrapped $D7$\ branes parameterized as
\begin{equation}
\Phi _{3,n}=z_{3}\ \Pi _{n},\qquad \Phi _{4,n}=z_{4}\ \Pi _{n},\qquad \Phi
_{5}=z_{5,n}\ \Pi _{n}\text{.}
\end{equation}
Therefore at the orbifold point of $\mathcal{R^{\prime }}/\mathbf{Z}_{5}$,
we have five fractional wrapped $D7$ branes. Moreover since near the
singularity, $\Phi _{1}$ and $\Phi _{2}$ split as
\begin{equation}
\Phi _{1}=\left( \lim_{z_{1}\rightarrow 0}\ {z_{1}}\right)
\sum_{n=1}^{5}a_{n}^{+},\qquad \Phi _{2}=\left( \lim_{z_{2}\rightarrow 0}\ {%
z_{2}}\right) \sum_{n=1}^{5}a_{n}^{-},
\end{equation}
there are also massless modes $\phi _{1,n}\sim a_{n}^{+}$\ and $\phi
_{2,n}\sim a_{n}^{-}$\ living on the wrapped $D7$ branes. They are
propagating modes traveling between $\Phi _{i,n}$\ and $\Phi _{i,n\pm 1}$\
fractional branes. The quiver diagram representing the fractional D branes
is also a $\Delta _{5}$\ pentagon with wrapped $D7$ siting at the vertices
and massless modes represented by the links. A similar analysis to that we
have developed above may be also written down for the $\mathbf{B}_{2}\bowtie
$\ $\mathcal{S}_{2}$\ and \ $\mathbf{B}_{3}\bowtie $\ $\mathcal{S}_{3}$
quintic fibrations. The five links joining the neighboring $\pi _{n}$\ nodes
describe massless fields of the effective field theory on the $D4$ branes at
singularity. For each fibration, there are $(2\times 5)$ massless complex
fields which we denote as $\chi _{a,k_{a}}\sim a_{k_{a}}^{+}$ and $\psi
_{a,k_{a}}\sim a_{k_{a}}^{-}$. In particular at $z_{1}=z_{2}=z_{3}=0$, we
have therefore triplets of massless fields as shown on table (\ref{tobib})
where we give the fields spectrum on the fractional wrapped $D7$ branes on $%
\mathbf{B}_{a}$ at the various singularities.

\textit{Fields Spectrum}\newline
\begin{equation}
\begin{tabular}{|c|c|c|}
\hline
Chiral fields $\rightarrow $ & Massless complex fields & Number of massless
fields \\[2mm] \hline
Singularity $\mathcal{S}_{1}$ & $\chi _{1,k_{1}}$ , $\psi _{1,k_{1}}$ & $%
2\times 5$ \\[2mm] \hline
Singularity $\mathcal{S}_{2}$ & $\chi _{2,k_{2}}$ , $\psi \ _{2,k_{2}}$ & $%
2\times 5$ \\[2mm] \hline
Singularity $\mathcal{S}_{3}$ & $\chi \ _{3,k_{3}}$ , $\psi \ _{3,k_{3}}$ & $%
2\times 5$ \\[2mm] \hline
Locus $z_{1}=0$ & $\chi _{1,k_{1}},\chi \ _{2,k_{2}},\chi \ _{3,k_{3}}$ & $%
3\times 5$ \\[2mm] \hline
\end{tabular}
\label{tobib}
\end{equation}
Such analysis can also be extended to the case where the orbifold is
realized as $\mathcal{Q}^{\left[ 2\right] }\sim \mathcal{R^{\prime \prime }}/%
\mathbf{Z}_{5}^{2}$.

\subsection{ Fractional Branes on $\mathcal{Q}^{\left[ 2\right] }$}

There are three manifest fibrations of the commutative quintic orbifold $%
\mathcal{Q}^{\left[ 2\right] }$ depending on the nature of the $\mathbf{Z}%
_{5}^{2}$ subgroups of $\mathbf{Z}_{5}^{3}$ one is considering. Let us
describe them briefly here below

\textbf{Fibration} $\mathcal{Q}^{\left[ 2\right] }\simeq CP^{1}\bowtie
\mathcal{S}_{12}$: \qquad Here the $\mathbf{Z}_{5}^{2}$ orbifold subsymmetry
of the $\mathbf{Z}_{5}^{2}$ group has the vector charges $q_{i}^{1}=\left(
1,-1,0,0,0\right) $ and $q_{i}^{2}=\left( 1,0,-1,0,0\right) $. The $%
q_{i}^{3} $ vector charge of the third factor may be thought as being set to
zero. As such the $\mathbf{B}_{12}$ basis of the fibration is parameterized
by the $z_{4}$ and $z_{5}$\ coordinates while the codimension one fiber $%
\mathcal{S}_{12}$ is given by the following complex surface with an $\mathbf{%
A}_{4}$ singularity,
\begin{equation}
u+v+t+\frac{t^{5}}{uv}=0.
\end{equation}
Like for the $\mathbf{B}_{a}\bowtie $\ $\mathcal{S}_{a}$\ fibrations, the\ $%
\mathbf{B}_{ab}$\ bases are parameterized by those holomorphic coordinates $%
z_{i}$ with $q_{i}^{a}=0$ and the $\mathcal{S}_{ab}$ fibers by the complex
variables with non zero $q_{i}^{a}$ charges with fixed points at the origin.

\textbf{Fibration} $\mathcal{Q}^{\left[ 2\right] }\simeq CP^{1}\bowtie
\mathcal{S}_{23}$:\qquad In this fibration the $\mathbf{Z}_{5}^{2}$ orbifold
subgroup has the following vector charges $q_{i}^{2}=\left(
1,0,-1,0,0\right) $ and $q_{i}^{3}=\left( 1,0,0,-1,0\right) $. The $\mathbf{B%
}_{23}$ basis and the $\mathcal{S}_{23}$\ fiber of the quintic are
respectively parameterized by $\left( z_{2},z_{5}\right) $\ and $\left(
z_{1},\ z_{3},\ z_{4}\right) $. Locally $\mathbf{B}_{23}\sim \mathbf{C}$
while the $\mathcal{S}_{23}$ singular surface with an $\mathbf{A}_{4}$
singularity at the origin.

\textbf{Fibration} $\mathcal{Q}^{\left[ 2\right] }\simeq CP^{1}\bowtie
\mathcal{S}_{31}$:\qquad In this case the CY vector charges of the $\mathbf{Z%
}_{5}^{2}$ orbifold subgroup are given by $q_{i}^{1}=\left(
1,-1,0,0,0\right) $ and $q_{i}^{3}=\left( 1,0,0,-1,0\right) $. The base is
parameterized by $z_{3}$ and $z_{5}$\ coordinates and the $\mathcal{S}_{31}$
fiber has a singularity at $z_{1}=z_{2}=z_{4}=0$. Its complex equation \ $%
u_{1}+u_{2}+t+\frac{t^{5}}{u_{1}u_{2}}=0$\ has an $\mathbf{A}_{4}$
singularity at $\ u_{1}=u_{2}=t=0$.

The main features of the various ${\mathbf{B}_{\left( ab\right) }}\bowtie
\mathcal{S}_{\left( ab\right) }$ fibrations are collected in the following
table

\begin{equation}
\begin{tabular}{|l|l|l|l|}
\hline
$F_{a}\otimes F_{b}$ & $F_{1}\otimes F_{2}$ & $F_{2}\otimes F_{3}$ & $%
F_{3}\otimes F_{1}$ \\ \hline
Fix Points & $\left( 0,0,0,z_{4},z_{5}\right) $ & $\left(
0,z_{2},0,0,z_{5}\right) $ & $\left( 0,0,z_{3},0,z_{5}\right) $ \\ \hline
$\mathbf{B}_{ab}$ & $\mathbf{CP}^{1}=\left\{ z_{4},z_{5}\right\} $ & $%
\mathbf{CP}^{1}=\left\{ z_{2},z_{5}\right\} $ & $\mathbf{CP}^{1}=\left\{
z_{3},z_{5}\right\} $ \\ \hline
\begin{tabular}{l}
Fibers \\
$\mathcal{S}_{ab}$%
\end{tabular}
&
\begin{tabular}{l}
$u_{1}+u_{2}+t+\frac{t^{5}}{u_{1}u_{2}}=0$ \\
$u_{1}=z_{1}^{5}$, $u_{2}=z_{2}^{5}$, \\
$t=z_{1}z_{2}z_{3}$.
\end{tabular}
&
\begin{tabular}{l}
$u_{1}+u_{3}+v+\frac{v^{5}}{u_{1}u_{2}}=0$ \\
$u_{1}=z_{1}^{5}$, $u_{3}=z_{3}^{5}$, \\
$v=z_{1}z_{3}z_{4}$%
\end{tabular}
&
\begin{tabular}{l}
$u_{1}+u_{4}+w+\frac{w^{5}}{u_{1}u_{4}}=0$ \\
$u_{1}=z_{1}^{5}$, $u_{4}=z_{4}^{5}$, \\
$w=z_{1}z_{2}z_{4}$.
\end{tabular}
\\ \hline
Fields & $\chi _{1,\left( k_{1},k_{2}\right) }$ $,$ $\psi _{1,\left(
k_{1},k_{2}\right) }$ & $\chi _{2,\left( k_{2},k_{3}\right) }$ $,$ $\psi
_{2,\left( k_{2},k_{3}\right) }$ & $\chi _{3,\left( k_{1},k_{3}\right) }$ $,$
$\psi _{3,\left( k_{1},k_{3}\right) }$ \\ \hline
\end{tabular}
\end{equation}
The singular representations of NC $\mathcal{Q}^{nc\left[ 2\right] }$ may be
obtained by starting from the regular representation eqs(\ref{salf}) and
taking the appropriate limits. For the $\mathcal{S}_{12}$ singular space,
the field moduli are associated with $q_{i}^{1}=q_{i}^{2}=0$ and taking$\
z_{1}$, $z_{2}$\ and $z_{3}$ to zero. We have for the$\mathcal{S}_{12}$
fiber,

\begin{eqnarray}
\Phi _{1} &=&\ \left( {{Q}\otimes }I_{id}\otimes I_{id}\right) \
\lim_{z_{1}\rightarrow 0}\ {z_{1}},\quad \Phi _{2}=\ \left( {{Q^{-1}}}%
\otimes I_{id}\otimes I_{id}\right) \ \lim_{z_{2}\rightarrow 0}\ {z_{2}},
\notag \\
\Phi _{3} &=&\ (I_{id}\text{ }{\otimes {Q^{-1}}\otimes }I_{id})\
\lim_{z_{3}\rightarrow 0}\ {z_{3}},\quad F_{1}=P\otimes Q^{\eta
_{12}}\otimes Q^{\eta _{13}},\quad F_{2}=Q^{\eta _{21}}\otimes P\otimes
Q^{\eta _{23}},
\end{eqnarray}
and for $\mathbf{B}_{12}$,

\begin{equation}
\Phi _{4}=z_{4}\ \otimes \mathbf{I}_{\mathcal{D}\left( G\right) },\quad \Phi
_{5}=z_{5}\ \otimes \mathbf{I}_{\mathcal{D}\left( G\right) },\quad
F_{3}=Q^{\eta _{31}}\otimes Q^{\eta _{32}}\otimes P.
\end{equation}
Once the limit to the singular point is taken, the non vanishing matrix
moduli $\Phi _{i}$,\ $i=4,5$,\ are proportional to the identity $\mathbf{I}_{%
\mathcal{D}\left( G\right) }$\ of\ $\mathcal{D}\left( G\right) $. As before
this property reflects just the existence of fractional $D$ branes at the
orbifold point $z_{1}=z_{2}=z_{3}=0$. To illustrate the idea, let us
reconsider the example of the $D9$ brane of type $IIB$ string wrapped on $%
\mathcal{Q}^{\left[ 2\right] }$; with local coordinates $\left( x^{\mu
};y_{1},...,y_{3}\right) $;. the $x^{\mu }$'s being the longitudinal non
compact coordinates of the $D3$ part of $D9$, while the $y$ compact
coordinates are as in eqs(\ref{rcor}). In presence of discrete torsion, the
wrapped $D9$ on\ $\mathcal{Q}^{\left[ 2\right] }$\ becomes a NC brane
generated by the algebra eqs(\ref{qint2}) where the $q_{i}^{a}$ charges are
as indicted above. At the fixed points of $\mathcal{Q}^{\left[ 2\right] }$
where a real four cycle shrinks to zero, the NC wrapped $D9$ give rise to
twenty five fractional wrapped $D5$ branes on $\mathbf{B}_{12}\sim \mathbf{CP%
}^{1}$. The transverse coordinates of these fractional $D5$ branes are given
by;

\begin{equation}
\Phi _{4;n,m}=x_{4}\ \Pi _{n,m},\quad \Phi _{5;n,m}=x_{5}\ \Pi _{n,m},
\end{equation}
where $\Pi _{n,m}$\ are the projectors on the $\mathcal{D}\left( \mathbf{Z}%
_{5}^{2}\right) $ representation states. The singular modes at the orbifold
point are carried by the ${Q\otimes Q}$, ${{Q^{-1}}}\otimes I_{id}$ and $%
I_{id}\otimes {{Q^{-1}}}$ operators of $\mathcal{D}\left( \mathbf{Z}%
_{5}^{2}\right) $. Moreover since near the singularity the $\Phi _{1}$, $%
\Phi _{2}$ and $\Phi _{3}$ operators may also be split as
\begin{equation}
\Phi _{1}=\lim_{z_{1}\rightarrow 0}\left( {x_{1}}\right)
\sum_{n_{1},n_{2}=1}^{5}A_{n_{1}}^{+}A_{n_{2}}^{+},\text{\quad }\Phi
_{2}=\lim_{z_{2}\rightarrow 0}\left( {x_{2}}\right)
\sum_{n_{1},n_{2}=1}^{5}A_{n_{1}}^{-}\Pi _{n_{2}},\quad \Phi
_{3}=\lim_{z_{3}\rightarrow 0}\left( {x_{3}}\right)
\sum_{n_{1},n_{2}=1}^{5}\Pi _{n_{1}}A_{n_{2}}^{-},
\end{equation}
one gets $3\times 25$ massless modes $\phi _{1;n_{1},n_{2}}\sim
A_{n_{1}}^{+}A_{n_{2}}^{+}$, $\phi _{2;n_{1},n_{2}}\sim A_{n_{1}}^{-}\Pi
_{n_{2}}$\ \ and $\phi _{2;n_{1},n_{2}}\sim \Pi _{n_{1}}A_{n_{2}}^{-}$\
living on the $D5$ branes wrapping $\mathbf{CP}^{1}$. They are propagators
between the $\Phi _{i;n_{1},n_{2}}$\ and $\Phi _{i,n_{1}\pm 1,n_{2}\pm 1}$\
fractional $D5$ branes. The quiver diagram representing the fractional $D5$
branes wrapping $\mathbf{CP}^{1}$ is a $\Delta _{5}\times \Delta _{5}$\
polygon with $D5$ branes sitting at the vertices and the $\phi
_{a;n_{1},n_{2}}$ massless modes propagating along the links; see figure 4.

\begin{figure}[tbh]
\begin{center}
\epsfxsize=10cm \epsffile{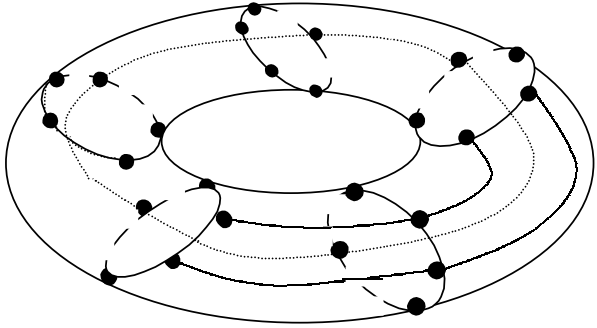}
\end{center}
\par
\caption{{\protect\small \textit{Here is represented the 25 vertices of
fractional $D$ branes quiver at $z_{1}=z_{2}=z_{3}=0$ singularity of $%
\mathbf{Z}^{2}$ orbifold subsymmetry. Links between vertices represent
massless modes of wrapped $D5$ branes on $\mathbf{CP}^{1}$ }}}
\end{figure}
An analogous analysis may be also made for the $\mathbf{B}_{23}\bowtie $\ $%
\mathcal{S}_{23}$\ and \ $\mathbf{B}_{31}\bowtie $\ $\mathcal{S}_{31}$
fibrations. As the results one gets are similar, we will omit this part. In
the end of this discussion, let us give the results for orbifold of $%
\mathcal{H}_{n}$.

\textbf{Extension:\qquad }Here we give the results for generic complex $n$
dimension hypersurface $z_{1}^{n+2}+...+z_{n+2}^{n+2}+a_{0}%
\prod_{i=1}^{n+2}z_{i}=0$ with orbifold group $\mathbf{Z}_{n+2}^{n}$. In
this case the torsion matrix $\eta _{ab}=\eta _{\left( ab\right) }+\eta _{%
\left[ ab\right] }$ has to belong to $SL\left( n,\mathbf{Z}\right) $ where
the antisymmetric part $\eta _{\left[ ab\right] }$ is non zero in presence
of discrete torsion. Geometrically, the corresponding NC geometry is an
elliptic fibration with base $\mathcal{H}_{n}$ and \ fiber a real $\left(
n+2\right) $ dimensional Fuzzy torus $\mathcal{T}_{\beta _{ij}}^{n+2}$. The $%
\beta _{ij}$ cocycles are as
\begin{equation}
\beta _{ij}=\exp i\frac{2\pi }{n+2}\eta _{\left[ ab\right]
}q_{i}^{a}q_{j}^{b}\text{,}
\end{equation}
where now the $n$ vectors $\mathbf{q}^{a}=\left( q_{i}^{a}\right) $\ have $%
\left( n+2\right) $ integer entries and satisfy the CY condition $%
\sum_{i=1}^{n+2}q_{i}^{a}=0$. These charge vectors can be chosen as,
\begin{eqnarray}
\mathbf{Z}_{n+2} &:&\qquad \mathbf{q}^{1}=\left( 1,-1,0,0,...,0,0,0\right)
,\qquad \qquad \qquad \qquad \left( 1\right)  \notag \\
\mathbf{Z}_{n+2} &:&\qquad \mathbf{q}^{2}=\left( 1,0,-1,0,...,0,0,0\right)
,\qquad \qquad \qquad \qquad \left( 2\right)  \notag \\
&&\qquad ...  \label{chn} \\
\mathbf{Z}_{n+2} &:&\qquad \mathbf{q}^{n-1}=\left( 1,0,0,0,...,-1,0,0\right)
,\qquad \qquad \qquad \left( n-1\right)  \notag \\
\mathbf{Z}_{n+2} &:&\qquad \mathbf{q}^{n}=\left( 1,0,0,0,...,0,-1,0\right)
.\qquad \qquad \qquad \qquad \left( n\right)  \notag
\end{eqnarray}
Fractional D-branes at orbifold points depend on the orbifold subgroups $G_{%
\left[ \alpha \right] }$ of $\mathbf{Z}_{n+2}^{n}$ one is considering.\
Since there are several subgroups in $\mathbf{Z}_{n+2}^{n}$, we have then
several possible representations. The natural ones are those associated with
the manifest factors $G_{\left[ \alpha \right] }=\mathbf{Z}_{n+2}^{\alpha }$
with $1\leq \alpha \leq \left( n-1\right) $. Let us give some comments
regarding this point, but forget for a while about string applications by
letting $n$ to be a generic positive integer greater than one and suppose
that we have a $p$-brane ( $p>n$ ) wrapping the $\mathcal{H}_{n}$\ compact
manifold.

(\textbf{a}) \textit{Fractional branes on} $\mathcal{H}_{n}/\mathbf{Z}_{n+2}$
\qquad Since there are $n$ manifest $\mathbf{Z}_{n+2}$\ subsymmetries in $%
\mathbf{Z}_{n+2}^{n}$, we have $n$ classes of fractional $\left( p-2\right) $%
-branes at the $n$ kinds of singularities $z_{1}=z_{a+1}=0$. Extending the
analysis we have made for the quintic to generic $\mathcal{H}_{n}$'s by
thinking about $\mathcal{H}_{n}/\mathbf{Z}_{n+2}$\ as a fiber bundle $%
\mathbf{B}_{a}\bowtie \mathcal{S}_{a}$ of base $\mathbf{CP}^{n-1}$ and a
fiber $\mathcal{S}_{a}$ given by the complex curve $u+v+\frac{v^{n+2}}{u}=0$
with an $\mathbf{A}_{n-1}$ singularity at the origin $u=v=0$, we will have $%
d_{1}$ fractional branes at each orbifold point ($d_{1}=\dim \mathcal{D}%
\left( \mathbf{Z}_{n+2}\right) $) and $2n$ massless modes living on the $%
\left( p-2\right) $ branes. Points in this NC geometry are represented by
polygons $\Delta _{n+2}$ with $\left( n+2\right) $\ vertices and $\left(
n+2\right) $ edges.

(\textbf{b}) \textit{Fractional branes on} $\mathcal{H}_{n}/\mathbf{Z}%
_{n+2}^{2}$ \qquad Before giving the result concerning $\mathcal{H}_{n}/%
\mathbf{Z}_{n+2}^{2}$, note that the action of $\mathbf{Z}_{n+2}^{2}$ on the
$z_{i}$ variables is additive. If one performs two successive $\mathbf{Z}%
_{n+2}$ actions with charges, say $q_{i}^{a}$ and $q_{i}^{b}$, then the
total $\mathbf{Z}_{n+2}^{2}$\ action has the charge $q_{i}^{a}+q_{i}^{b}$,
\begin{eqnarray}
\mathbf{Z}_{n+2} &:&\qquad z_{i}\longrightarrow z_{i}\omega
^{q_{i}^{a}},\quad \mathbf{Z}_{n+2}:\qquad z_{i}\longrightarrow z_{i}\omega
^{q_{i}^{b}}, \\
\mathbf{Z}_{n+2}^{2} &:&\qquad z_{i}\longrightarrow z_{i}\omega
^{q_{i}^{a}+q_{i}^{b}}.  \notag
\end{eqnarray}
From these relations, one sees that this $\mathbf{Z}_{n+2}^{2}$ action is
equivalent to a $\mathbf{Z}_{n+2}^{\prime }$ diagonal subsymmetry with $%
q_{i}^{a}+q_{i}^{b}$ charge. The remaining $q_{i}^{a}-q_{i}^{b}$ charge is
associated with the off diagonal subsymmetry which play no role here. Using
eqs(\ref{chn}), we get,
\begin{equation}
\mathbf{Z}_{n+2}^{2}:\qquad \mathbf{q}^{1}+\mathbf{q}^{2}=\left(
2,-1,-1,0,...,0,0,0\right) .  \label{prim}
\end{equation}
The fixed points of (\ref{prim}) are at $z_{1}=z_{2}=z_{3}=0$; but for
generic $q_{i}^{a}+q_{i}^{b}$ vector charge fixed points are at $%
z_{1}=z_{a+1}=z_{b+1}=0$. This property shows that the $\mathcal{H}_{n}^{nc}$
geometries have the following features: (i) the $\mathbf{B}_{ab}\bowtie
\mathcal{S}_{ab}$ realizations of $\mathcal{H}_{n}^{nc}$ have a $\mathbf{CP}%
^{n-2}$ base and a fiber $\mathcal{S}_{ab}$ described by the complex surface
$u+v+t+\frac{t^{n+2}}{uv}=0$ with an $\mathbf{A}_{n-1}$ singularity at the
origin $u=v=t=0$. (ii) there are $\frac{n\left( n-1\right) }{2}$ possible
fibrations for $\mathcal{H}_{n}/\mathbf{Z}_{n+2}^{2}$ and $\frac{n\left(
n-1\right) }{2}$\ classes of fractional $\left( p-4\right) $-branes at the $%
z_{1}=z_{a+1}=z_{b+1}=0$ singularities. Each class contains $d_{1}d_{2}$
fractional $D\left( p-4\right) $-branes and $2\left(
d_{1}d_{2}-d_{1}-d_{2}\right) $\ massless modes living on theses branes ($%
d_{i}=\dim \mathcal{D}_{i}\left( \mathbf{Z}_{n+2}\right) $). Points in this
NC geometry are given by the crossed product of two polygons $\Delta
_{d_{1}}\times \Delta _{d_{2}}$.

(\textbf{c}) \textit{Fractional branes on} $\mathcal{H}_{n}/\mathbf{Z}%
_{n+2}^{k}$ \qquad Extending the above reasoning to $\mathcal{H}_{n}/\mathbf{%
Z}_{n+2}^{k}$, the orbifold symmetry $\mathbf{Z}_{n+2}^{k}$\ with charge
vectors $\mathbf{q}^{a_{1}}$, $...$, $\mathbf{q}^{a_{k}}$,\ acts in practice
through ist diagonal $\mathbf{Z}_{n+2}$ subgroup with a CY charge $Q_{\left(
a_{1},...,a_{k}\right) }=\mathbf{q}^{a_{1}}+...+\mathbf{q}^{a_{k}}$; it has
fixed points located at $z_{1}=z_{a_{1}+1}=...=z_{a_{k}+1}=0$. For the
example of the leading $k$ factors $\mathbf{Z}_{n+2}$ of $\mathbf{Z}%
_{n+2}^{n}$. of eqs(\ref{chn}), the vector charge $Q_{\left( 1,...,k\right)
} $ of $\mathbf{Z}_{n+2}^{k}$ is given by;
\begin{equation}
\mathbf{Z}_{n+2}^{k}:\qquad Q_{\left( 1,...,k\right) }=\left(
k,-1,-1,...,-1,0,...,0,0\right) ;
\end{equation}
where the first $\left( k+1\right) $ entries are non zero and all remaining
ones are null. A simple counting of the degrees of freedom shows that there
are $\frac{n!}{\left( n-k\right) !k!}$ possible $\mathbf{Z}_{n+2}^{k}$
subgroups in $\mathbf{Z}_{n+2}^{n}$. The $\mathbf{B}_{\left(
a_{1}...a_{k}\right) }\bowtie \mathcal{S}_{\left( a_{1}...a_{k}\right) }$
representations of the manifold $\mathcal{H}_{n}/\mathbf{Z}_{n+2}^{k}$ have
a $\mathbf{CP}^{n-k}$ base and a fiber $\mathcal{S}_{\left(
a_{1}...a_{k}\right) }$ given by the complex dimension $k$ hypersurface $%
u_{1}+...+u_{k}+t+\frac{t^{n+2}}{\varrho }=0$ with $\varrho
=\prod_{j=1}^{k}u_{j}$ and an $\mathbf{A}_{n-1}$ singularity at $%
u_{1}=...=u_{k}=t=0$.\ Together with these realizations, there are $\frac{n!%
}{\left( n-k\right) !k!}$\ classes of fractional $D\left( p-2k\right) $%
-branes at the $u_{1}=...=u_{k}=t=0$ singularities. Each class contains $%
\prod_{j=1}^{k}d_{j}$ fractional $\left( p-2k\right) $-branes and $%
k\prod_{j=1}^{k}d_{j}-\sum_{i=1}^{k}\prod_{j\neq i}d_{i}$\ \ massless modes
living on theses.

\section{Conclusion}

Using results on type II string compactification on CY orbifolds and the
algebraic geometry method of Berenstein and Leigh, we have have developed\
the study fractional $D$ branes on generic complex $n$ dimension NC
orbifolds of CY hypersurfaces $\mathcal{H}_{n}$. This is an explicit study
which give the general solutions for NC geometry and complete by the way
special results obtained previously in \cite{2,24}. It also allows a more
insight in NC geometry induced by discrete torsion and recovers as a
particular case stringy constructions in connection with orbifolds of $D=4$
SYM theory embedded in type II string compactifications on local CY
orbifolds with discrete torsion \cite{38}. The general solutions we have
obtained have geometric and algebraic interpretations and can moreover be
viewed as an explicit verification of Adams and Fabinger conjecture
regarding emergent dimensions considered recently in \cite{46}.

Geometrically, we have shown that the NC orbifolds with discrete torsion are
special elliptic fiber bundles based on $\mathcal{H}_{n}$ and fuzzy torii as
fibers. The latters resolve singularities and lead to a fractionalisation of
D branes due to the complete reducibility property of the representations of
the orbifold group at singularities. In the large $n$ limit our analysis
extends naturally; the discrete $\mathbf{Z}_{n+2}$ group factors tends to $%
\mathbf{U}\left( 1\right) $ and the original $Dp$ branes are mapped to $Dp+2$
ones in accord with the prediction of \cite{46}; but also with the explicit
computation made in \cite{2,24}. In the continuous limit the solutions are
quite similar; rational torii fibers considered in the present study are
replaced in the continuum by irrational ones; for details see \cite{24,33}.

Algebraically, the general solutions we have derived in this paper offers,
amongst others, a remarkable classification of NC orbifolds. This
classification is completely characterized by the two following: (i) the $%
q_{i}^{a}$ vector charges of the orbifold group $\mathbf{Z}_{n+2}^{n}$ with $%
n>1$ and (ii) a $n\times n$ matrix $\eta _{ab}=\eta _{\left( ab\right)
}+\eta _{\left[ ab\right] }$ of the group SL$\left( n,Z\right) $ defining
the various possible orbifolds. Discrete torsion exists whenever the $\eta _{%
\left[ ab\right] }$ antisymmetric part is non zero. In addition to the fact
that they go beyond the known ones in literature, our solutions exhibit
manifestly the\ discrete torsion dependence embodied by $\eta _{\left[ ab%
\right] }$ and\ full orbifold geometric invariance carried by the $q_{i}^{a}$%
s. In this regards it is worthwhile to recall that the $\beta _{ij}$\
cocycles appearing in $\ Z_{i}Z_{j}=\beta _{ij}Z_{j}Z_{i}$ NC geometry
relations read, in terms of $q_{i}^{a}$ charges and the matrix $\eta _{ab}$,
as $\beta _{ij}=\exp i\frac{2\pi }{n+2}\eta _{\left[ ab\right]
}q_{i}^{a}q_{j}^{b}$.

Supersymmetric field theoretic aspects of this construction seems to be
linked to deformations by $\mathcal{N}=1$ adjoint matter superpotentials;
progress in this issue will be considered in a future occasion.

\begin{acknowledgement}
I am grateful to Julius Wess for kind hospitality at Munich University where
part of this work has been done. I thank A.Belhaj, M.Bennai and E.M Sahraoui
for earlier collaborations on this subject. This work is supported by
Protars III, CNRST, Rabat, Morocco.
\end{acknowledgement}


\begin{thebibliography}{99}
\bibitem{1}  David Berenstein, Vishnu Jejjala, Robert G. Leigh, D-branes on
Singularities: New Quivers from Old,Phys.Rev. D64 (2001) 046011,
hep-th/0012050.

\bibitem{2}  David Berenstein, Robert G.Leigh, Phys.Lett. B499
(2001),hep-th/0009209.

\bibitem{3}  Matthias Klein, Raul Rabadan, $Z_{N }x Z_{M} $ orientifolds
with and without discrete torsion, JHEP 0010 (2000) 049, hep-th/0008173.

\bibitem{4}  Eric R. Sharpe, Recent Developments in Discrete Torsion,
Phys.Lett. B498 (2001) 104-110, hep-th/0008191.

\bibitem{5}  Paul S. Aspinwall, M. Ronen Plesser, D-branes, Discrete Torsion
and the McKay Correspondence, JHEP 0102 (2001) 009, hep-th/0009042.

\bibitem{6}  Paul S. Aspinwall, A Note on the Equivalence of Vafa's and
Douglas's Picture of Discrete Torsion, JHEP 0012 (2000) 029, hep-th/0009045.

\bibitem{7}  Keshav Dasgupta, Seungjoon Hyun, Kyungho Oh, Radu Tatar,
Conifolds with Discrete Torsion and NCG, JHEP 0009 (2000) 043,
hep-th/0008091.

\bibitem{8}  M.R Douglas, D-branes and Discrete torsion, hep-th/9807235.

\bibitem{9}  M.R Douglas and B.Fiol, D-branes and Discrete torsion,
hep-th/9903031.

\bibitem{10}  D.Berenstein, R.G.Leigh, Discrete torsion and Duality, JHEP 01
(2000) 038, hep-th/0001055.

\bibitem{11}  A. Connes, M.R. Douglas et A. Schwarz, JHEP 9802, 003 (1998),
hep-th/9711162.

\bibitem{12}  N. Seiberg and E. Witten, JHEP 9909(1999) 032, hep-th/990814.

\bibitem{13}  David J. Gross, Nikita A. Nekrasov, Solitons in NC Gauge
Theory, 0103 (2001) 044 hep-th/0010090.

\bibitem{14}  I. Benkaddour , M. Bennai, E. Diaf and H. Saidi CQG {17}%
(2000)1765.

\bibitem{15}  A. Belhaj, M. Hssaini, E.M.Sahraoui, E. H.
Saidi,Class.Quant.Grav.18(2001) 2339, hep-th/0007137.

\bibitem{16}  N. Nekrasov and A. Schwarz, Commun Math. Phys \textbf{198}%
(1998) 689-703, hep-th/9802068.

\bibitem{17}  D.Berenstein, R.G. Leigh, Resolution of Stringy Singularities
by NCAs, JHEP 0106 (2001) 030, hep-th/0105229.

\bibitem{18}  M. Bertolini, P. Di Vecchia, M. Frau, A. Lerda, R. Marotta,
N=2 Gauge theories on systems of fractional D3/D7 branes, Nucl.Phys. B621
(2002) 157-178, hep-th/010705.

\bibitem{19}  M. Cvetic, G.W. Gibbons, James T. Liu, H. Lu, C.N. Pope, A New
Fractional D2-brane, $G_2$ Holonomy and T-duality, hep-th/010616.

\bibitem{20}  Tadashi Takayanagi, Holomorphic Tachyons and Fractional
D-branes, Nucl.Phys. B603 (2001) 259-285, hep-th/0103021.

\bibitem{21}  Subir Mukhopadhyay, Koushik Ray, Fractional Branes on a
Non-compact Orbifold, JHEP 0107 (2001) 007, hep-th/0102146.

\bibitem{22}  M. Frau, A. Liccardo, R. Musto, The Geometry of Fractional
Branes, Nucl.Phys. B602 (2001) 39-60, hep-th/0012035.

\bibitem{23}  M. Billo', L. Gallot, A. Liccardo,Fractional branes on ALE
orbifolds , the proceedings of the RTN meeting ``The Quantum Structure of
Spacetime and the Geometric Nature of Fundamental Interactions'', (Corfu,
September 2001) , hep-th/0112190 .

\bibitem{24}  A. Belhaj, E.H. Saidi, On NCCY Hypersurfaces, hep-th/0108143,
Phys.Lett. B523 (2001) 191-198.

\bibitem{25}  I. Bars, H. Kajiura, Y. Matsuo, T. Takayanagi,Tachyon
Condensation on NC Torus,Phys.Rev. D63 (2001) 086001, hep-th/0010101.

\bibitem{26}  B. R. Greene, C. I. Lazaroiu, P.Yi, D-particles on $%
T^{4}/Z_{n} $ orbifolds, hep-th/9807040; Nucl.Phys.B539 (1999) 135.

\bibitem{27}  E. M. Sahraoui, E.H. Saidi, Solitons in large NC
Class.Quant.Grav.18 (2001) 3339, hep-th/0012259,

\bibitem{28}  E. M. Sahraoui, E.H. Saidi, D-branes on NC Orbifolds, JHEP
0205 (2002) 063, hep-th/0105188.

\bibitem{29}  Julius Wess and J Bagger, Supersymmetry and Supergravity,
Princeton Univ. Press, (1983).

\bibitem{30}  Juan M. Maldacena, The Large N Limit of Superconformal Field
Theories and Supergravity, Adv.Theor.Math.Phys. 2 (1998) 231-252;
Int.J.Theor.Phys. 38 (1999) 1113-1133 , hep-th/9711200.

\bibitem{31}  E.Witten, Multi-Trace Operators, AdS/CFT Correspondence,
hep-th/0112258.

\bibitem{32}  O. Aharony, S.S. Gubser, J. Maldacena, H. Ooguri, Y. Oz, Large
N Field Theories, String Theory and Gravity, Phys.Rept. 323 (2000) 183-386
,hep-th/9905111 .

\bibitem{33}  M. Bennai, E.H. Saidi, NCCY in Toric Varieties with NC
fibration, Phys.Lett. B550 (2002) 108-116.

\bibitem{34}  S. Gukov, C. Vafa, E. Witten, CFT's From CY4s, Nucl.Phys. B584
(2000) 69-108, hep-th/9906070.

\bibitem{35}  K.Hori, H.Ooguri, C.Vafa, Non-Abelian Conifold Transitions and
N=4 Dualities in 3D, Nucl.Phys. B504 (1997) 147-174, hep-th/9705220.

\bibitem{36}  Julius Wess, Proc of Workshop on High Energy Physics 2
(2000)1-11. Rabat University.

\bibitem{37}  Philip Candelas, Eugene Perevalov, Govindan Rajesh, Toric
Geometry and Enhanced Gauge Symmetry of F-Theory/Heterotic Vacua, Nucl.Phys.
B507 (1997) 445-474, hep-th/9704097.

\bibitem{38}  A. Lawrence, N. Nekrasov, C. Vafa, On CFT$_{4}$, Nucl.Phys.
B533 (1998) 199-209, hep-th/9803015.

\bibitem{39}  A. Belhaj, E.H.Saidi, Hyperkahler Singularities in
Superstrings Compactification and N=4 CFT$_{2}$, Class.Quant.Grav. 18 (2001)
57-82,hep-th/0002205.

\bibitem{40}  B. Jurco, L.M\"{o}ller, S. Schraml, P. Schupp, J. Wess,
Eur.Phys.J. C21 (2001) 383-388, hep-th/0104153.

\bibitem{41}  J. Madore, S. Schraml, P. Schupp, J. Wess, Gauge Theory on NC
Spaces, Eur.Phys.J. C16 (2000) 161-167, hep-th/0001203.

\bibitem{42}  S.Katz, D.R. Morrison, M. R. Plesser, Enhanced Gauge Symmetry
in Type II String Theory, Nucl.Phys. B477 (1996) 105-140, hep-th/9601108.

\bibitem{43}  S.Katz, P.Mayr, C.Vafa, Mirror symmetry and Exact Solution of
4D N=2 Gauge Theories I, Adv.Theor.Math.Phys. 1 (1998) 53-114,
Adv.Theor.Math.Phys. 1 (1998) 53-114.

\bibitem{44}  N.C.Leung, C.Vafa, Branes and Toric Geometry,
Adv.Theor.Math.Phys.2 (1998) 91-118, hep-th/9711013.

\bibitem{45}  A. Belhaj, A. E. Fallah , E. H, Saidi, CQG 17 (2000)515-532.

\bibitem{46}  A. Adams, M. Fabinger, Deconstructing NC with a Giant Fuzzy
Moose, JHEP 0204 (2002) 006.

\bibitem{47}  Nima Arkani-Hamed, Andrew G. Cohen, Howard Georgi,
(De)Constructing Dimensions, Phys.Rev.Lett. 86 (2001) 4757-4761.

\bibitem{48}  A.El Rhalami, E.M.Sahraoui, E.H.Saidi, NC Branes,Hierarchies
in QHFs, JHEP 0205:004,2002.

\bibitem{49}  A.El Rhalami, E.H. Saidi, NC Effective gauge model for FQH
states, JHEP 0210:039,2002.
\end{thebibliography}
\end{document}